\documentclass[11pt,a4paper,oneside]{article}
\usepackage[top=3cm, bottom=3cm, left=2cm, right=2cm]{geometry}
\usepackage[english]{babel}
\usepackage[utf8]{inputenc}
\usepackage{enumerate}
\usepackage[T1]{fontenc}
\usepackage{setspace}
\usepackage{xcolor}
\usepackage{booktabs}
\usepackage{multicol}
\usepackage{multirow}
\usepackage{makecell}
\usepackage{subfigure}
\usepackage{xr}
\usepackage{algorithm}
\usepackage{algpseudocode}
\usepackage{subcaption}
\usepackage{tikz}
\usepackage{blkarray}
\usetikzlibrary{positioning}
\usetikzlibrary{fit}
\usetikzlibrary{arrows}
\usetikzlibrary{graphs}
\usetikzlibrary{matrix}
\usepackage{amsthm,amsmath,amssymb,mathrsfs,dsfont,mathtools, mathabx}
\usepackage{bbm}
\usepackage{bm}
\usepackage{graphicx}
\usepackage[colorlinks=true, allcolors=blue]{hyperref}
\usepackage{comment}
\usepackage[all]{nowidow}

\usepackage[authoryear, round]{natbib}
\allowdisplaybreaks
\usepackage{authblk}

\date{ \vspace{-5mm} }
\title{Learning discrete Bayesian networks with hierarchical Dirichlet shrinkage}
\author[1]{Alexander Dombowsky}
\author[2,3]{David B. Dunson}
\affil[1]{Gladstone Institute of Data Science and Biotechnology, San Francisco, CA, USA}
\affil[2]{Department of Statistical Science, Duke University, Durham, NC, USA}
\affil[3]{Department of Mathematics, Duke University, Durham, NC, USA}

\usepackage[sf]{titlesec}
\usepackage{sectsty}
\allsectionsfont{\centering \normalfont\scshape} % Make all sections centered, the default font and small caps

\theoremstyle{definition}

\newtheorem{theorem}{Theorem}

\newtheorem{prop}{Proposition}

\def\bs{\boldsymbol}
\def\lb{\left\{ }
\def\rb{\right\}}
\def\R{\mathbb{R}}
\def\E{\mathbb{E}}

\def\diff{\tx{d} \:}

\def\tx{\textrm}

\def\J{\mathcal J}

\def\Ca{\tx{Ca}}
\def\xnew{\bs x^{\tx{new}}}

\def\btildepi{\Tilde{\bs \pi}}
\def\bomega{\bs \omega}

\def\N{\mathcal{N}}

\def\Pa{\tx{Pa}}

\newcommand{\ind}{\perp\!\!\!\!\perp}

\begin{document}

\maketitle

\begin{abstract}

A discrete Bayesian network (DBN) is a directed acyclic graph (DAG) consisting of categorical variables. Two popular approaches for DBN modeling include classification and nonparametric methods. However, both methods often require a large number of parameters, such as high-order interactions in the former and cell probabilities in the latter. In this article, we propose a hierarchical model for node-parent conditional probabilities, inducing shrinkage to low-dimensional latent parameters aposteriori. We generate samples from the posterior distribution of these latent variables using the Metropolis-adjusted Langevin algorithm (MALA) within a Gibbs sampler. Moreover, we verify that the full conditional distribution is log-concave under mild conditions, facilitating efficient sampling. We then detail several algorithms for structure learning that incorporate our hierarchical prior and preserve the DAG property. Through simulations, we evaluate the performance of our method for sparse counts, discovering graph structure, and selecting between competing DAGs.  We conclude with an application to uncovering prognostic network structure from a breast cancer dataset. 
\end{abstract}
{\textit{Keywords}: Categorical data; directed acyclic graphs; Dirichlet distribution; structure learning; variable selection}
\doublespacing

\newpage

\section{Introduction}  \label{section:intro} 

Graphical models convey the conditional independencies in a group of random variables and are used routinely in many fields. The focus of this article is on directed acyclic graphs (DAGs) for categorical observed variables, which are
also known as belief networks, a type of \textit{Bayesian network}
\citep{pearl1985bayesian, pearl1988probabilistic}. Such models express the conditional independencies in random variables $\bs x = (x_1, \dots, x_p)$ in terms of a DAG $G = (V,E)$. The DAG has nodes corresponding to the $p$ variables, $V = [p] = \lb 1, \dots, p \rb$ for $j \in [p]$. A DAG can equivalently be expressed in terms of the parent sets of each node, or $\Pa(j) = \lb j^\prime<j : (j^\prime, j) \in E \rb$. By definition, we require $x_j \ind \lb \bs x_{[j-1] \setminus \Pa(j)} \rb \mid \bs x_{\Pa(j)}$, where for any $U \subset V$, we set $\bs x_U = (x_{j^\prime})_{j^\prime \in U}$. When $x_j \in [k_j]$ for all $j \in [p]$, conditional independence is conveyed mathematically through a probability mass function (PMF) $\Pr(\bs x)$. The resulting probabilistic model is referred to as a discrete Bayesian network (DBN). 

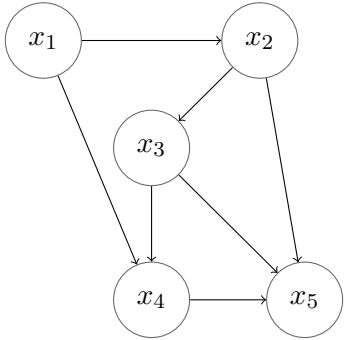
\begin{figure}[t]
    \centering
\begin{tikzpicture}[roundnode/.style={circle, draw=black!60, minimum size=10mm}, squarednode/.style={rectangle, draw=red!60, fill=red!5, thick, minimum size=5mm},box/.style = {draw,black,inner sep=10pt,rounded corners=5pt}, node distance = {10mm}]
        % nodes
        \node[roundnode] (3) {$x_3$};
        \node[roundnode] (1) [above left = of 3] {$x_1$};
        \node[roundnode] (2) [above right = of 3] {$x_2$};
        \node[roundnode] (4) [below = of 3] {$x_4$};
        \node[roundnode] (5) [right = of 4] {$x_5$};
        % arrows
        \draw[->] (1)--(2);
        \draw[->] (1)--(4);
        \draw[->] (2)--(3);
        \draw[->] (3)--(4);
        \draw[->] (2)--(5);
        \draw[->] (3)--(5);
        \draw[->] (4)--(5);
        % % boxes
    \end{tikzpicture}
    \caption{An example DAG for variables $x_1$, $x_2$, $x_3$, $x_4$, and $x_5$.}
    \label{fig:example-dag}
\end{figure}

The example DAG in Figure \ref{fig:example-dag} factorizes as $\Pr(\bs x) = \Pr(x_1) \Pr(x_2 \mid x_1) \Pr(x_3 \mid x_2) \Pr(x_4 \mid x_1, x_3) \Pr(x_5 \mid x_2,x_3,x_4)$. Each term in the factorization only depends on the nodes in its parent set. We can extract further information on the conditional dependence relations between $x_j$ and $\bs x_{[p] \setminus \lb j \rb}$ using a Markov blanket of node $j$, which is defined as the union of $\Pa(j)$, its children $\lb j^\prime \in [p]: (j,j^\prime) \in E \rb$, and the parent set of its children \citep{pearl1988probabilistic}. In practice, DBNs are applied for (a) training a predictive model for a known DAG, and (b) inferring the DAG itself, which is known as \textit{structure learning} \citep{kitson2023survey}. We focus on both tasks in this article. 

Contemporary Bayesian models for DBNs often use multinomial likelihoods with conjugate Dirichlet priors on the conditional probabilities \citep{bishop2006pattern, scutari2021bayesian}. 
%A posteriori, the conditional probabilities of a node given its parent set are also Dirichlet-distributed, with a high-dimensional array of concentration hyperparameters updated by the conditional cell counts. 
A drawback of the Dirichlet-multinomial method is the dimensionality. For instance, in binary data, there are $2^{|\Pa(j)|}$ parameters, which are the success probabilities for each combination of categories in the parent set. Also, cell counts are often sparse, so the hyperparameters of the Dirichlet prior have a substantial impact on the posterior. There is little practical guidance on how to choose these hyperparameters despite their influence. Alternately, one can use a standard classification and function approximation algorithms to build a predictive model given a DAG \citep{rijmen2008bayesian,kratzer2023additive}, or to infer the structure of the network, but these often require modeling high-order interaction terms, since all of the predictors are categorical.

In this paper, we propose a flexible Bayesian method for DBN classification and structure learning, known as Hierarchical Directed Dirichlet Networks (HiDDeN). Conditional probabilities are modeled hierarchically across the parent categories. After marginalizing out the high-dimensional cell probabilities, we obtain a low-dimensional latent representation of our model, which has $k_j$ parameters, or \textit{concentrations}, for each node. We obtain a Metropolis-within-Gibbs sampler to simulate from the posterior of the concentrations, which uses the Metropolis-adjusted Langevin algorithm (MALA). Interestingly, these strategies are applicable beyond DBNs; for example, for models built from
hierarchical Dirichlet processes (HDP) \citep{teh2006sharing}. We show that the full conditional distribution of the concentrations factorizes across the categories of each node and, under certain conditions on our hyperparameters and if the marginal cell counts are positive, is almost surely log-concave.

We extend HiDDeN to account for the uncertainty in the DAG, allowing inference on $\Pr(G \mid \bs n)$ and applicability to causal discovery \citep{pearl2009causality}. We initially focus on the problem of inferring a single parent set. The first algorithm we propose assumes all possible parent sets have positive probability and iteratively adds and removes edges. A second algorithm is detailed for the case in which the user would like to evaluate a moderate collection of candidate parents. Both retain the DAG structure and allow for the incorporation of prior knowledge and regularization (i.e., restricting the degree of a node). We show that these algorithms can also be used to model Markov blankets, then modify them to produce MCMC samples of DAGs, allowing for estimation and uncertainty quantification of $G$.

%The output is used to obtain a point estimate of $G$ and express uncertainty with posterior edge probabilities $\Pr((j^\prime,j) \mid \bs n)$. 

There is a rich literature on hierarchical Dirichlet models and their uses. A notable contribution is the HDP \citep{teh2006sharing}, which was originally motivated by modeling dependence in group-specific distributions and clusterings. Hierarchical Dirichlet distributions are also widely used in topic modeling. For example, \cite{adler2011hierarchically} extend latent Dirichlet allocation \citep{blei2003latent} to shrink towards a common Dirichlet measure across a corpus of documents. However, none of these methods are intended for modeling probability distributions that factorize according to a DAG, nor do they address structure learning. Other works propose versions of the HDP that incorporate a DAG structure \citep{zhang2010evolutionary, alam2019tree, chakrabarti2024graphical}, although these models differ greatly from ours.

\cite{azzimonti2017hierarchical} and \cite{azzimonti2019hierarchical} apply a finite version of the HDP to DBNs, while \cite{azzimonti2022bayesian} develop a graph score for related models. Their methodology is extended to account for multi-source data in \cite{cerezo2023parametric}. These works focus on meta-analysis and require that the sum of the concentrations, denoted $\beta_j$, is prespecified. We allow for $\beta_j$ to be learned from the data aposteriori. Additionally, these works use variational methods to estimate the concentration parameters and use a variational surrogate in place of the marginal likelihood; such variational approaches are subject to approximation error to the true posterior in practice. 
In contrast, we make major advances in the ability to conduct accurate Bayesian inferences by developing efficient posterior sampling using MCMC. Furthermore, HiDDeN enables structure learning through fully Bayesian models, whereas the above approaches rely on search algorithms over the DAG space. There is also a literature on recursive Dirichlet priors for tree-structured DAGs \citep{petitjean2018accurate, zhang2020bayesian}, though these incorporate a markedly different structure than the hierarchical Dirichlet model we propose.

The remainder of this paper is organized as follows. Section \ref{section:methods} details the graph-theoretic and contingency table notation, motivates hierarchical modeling for DBNs, and introduces HiDDeN. A MALA-within-Gibbs sampler for HiDDeN is described in Section \ref{section:posterior}, and we propose several algorithms for structure learning in Section \ref{section:structure-learning}. Section \ref{section:simulations} evaluates HiDDeN's performance for parameter and structure learning on synthetic data, while Section \ref{sect:METABRIC} uses HiDDeN to uncover network structures related to breast cancer treatment. Methodological and theoretical extensions are discussed in Section \ref{section:discussion}.

\section{Constructing Dirichlet Networks for Known DAGs} \label{section:methods}

\subsection{Discrete Bayesian Networks}

Let $\bs x = (x_1, \dots, x_p)$ be a vector of $p$ variables, where $x_j \in [k_j] = \lb 1, \dots, k_j \rb$ with $2 \leq k_j<\infty$ for all $j \in [p] = \lb 1, \dots, p\rb$. If $k_j=2$, $x_j$ is binary and we follow the convention $x_j \in \lb 0,1 \rb$. The vector $\bs x$ takes values in $\chi = \lb (x_1, \dots, x_p) : x_j \in [k_j] \rb$. We are interested in modeling the joint probability mass function $\Pr(\bs x)$, which corresponds to an unknown probability tensor of dimension $K = \prod_{j=1}^p k_j$. In general, the massive dimension of the probability tensor is unwieldy, motivating the more interpretable approach of describing $\Pr(\bs x)$ through its \textit{conditional independencies}. We say that $x_{j_1}$ and $x_{j_2}$ are conditionally independent given a set of variables $\bs x_{j_3}$ if $\Pr(x_{j_1},x_{j_2}\mid \bs x_{j_3}) = \Pr(x_{j_1} \mid \bs x_{j_3} )\Pr(x_{j_2} \mid \bs x_{j_3})$ for all choices of $x_{j_1}, x_{j_2}, \bs x_{j_3}$. 

A \textit{Bayesian network} is a framework for describing the conditional dependence structure between variables comprising $\bs x$, characterized by a directed acyclic graph (DAG) $G = ([p], E)$ and parameters $\bs \pi$ \citep{pearl1985bayesian,pearl1988probabilistic}. The graph $G$ has the vertex set $V = [p] = \lb 1, \dots, p \rb$ and edges $E \subseteq \lb (j,j^\prime) : j \neq j^\prime \in [p] \rb$. A DAG is a specific type of graph that has (a) directed edges and (b) no cycles (i.e., a non-trivial path in $G$ from a node to itself).
A discrete Bayesian network (DBN) models the joint distribution of a set of categorical variables $\bs x$ as
\begin{equation} \label{eq:Bayesian-network}
\begin{gathered}
    \Pr(\bs x \mid \bs \pi, G) =  \prod_{j=1}^p \Pr(x_j \mid \bs x_{\Pa(j)}, G) 
    = \prod_{j=1}^p \bs \pi_{j \mid \Pa(j)}(x_j \mid \bs x_{\Pa(j)}) \tx{ for all } \bs x \in \chi,
\end{gathered}
\end{equation}
where $\Pa(j) = \lb j^\prime < j: (j^\prime, j) \in E \rb$ indexes the \textit{parents} of $j$, $\bs x_{\Pa(j)} = (x_{j^\prime})_{j^\prime \in \Pa(j)}$ is the vector of variables with nodes in $\Pa(j)$, and $\bs \pi_{j \mid \Pa(j)}(x_j \mid \bs x_{\Pa(j)}) = \Pr(x_j \mid \bs x_{\Pa(j)}, G)$. We use $E_j = \lb (j^\prime,j) \in E: j^\prime <j \rb$ to represent the edges associated with the parents of node $j$. By convention, any node with no parents is a \textit{root node}, and if node $j$ is a root, this implies $\bs \pi_{j \mid \Pa(j)}(x_j \mid \bs x_{\Pa(j)}) = \bs \pi_j(x_j)$.

DBNs essentially describe a connection between a graph $G$ and the joint probability of a set of variables. Formally, we are assuming a relationship between the conditional independencies in $\bs x$ and the edges in $G$. This requires that $G$ be an \textit{independence map} (I-map) \citep{pearl1988probabilistic}, i.e., that if the nodes corresponding to $j_3$ intercept all directed paths between $j_1$ and $j_2$, then $x_{j_1} \ind x_{j_2} \mid \bs x_{j_3}$. To ensure consistency between $G$ and \eqref{eq:Bayesian-network}, we further assume that $G$ is a \textit{minimal I-map}; if $H \subset G$ is the subgraph of $G$ defined by deleting any edge in $E$, then $H$ is not an I-map for $\bs x$.

\subsection{Modeling of the Cell Counts}

Let $\bs x_{i} = (x_{i1}, \dots, x_{ip})$ for $i=1, \dots, n$ be categorical observations, i.e. $x_{ij} \in [k_j]$ for all $j=1, \dots, p$, which we collect in $\lb \bs x_i \rb$. For a multinomial model, the sufficient statistics are the \textit{cell counts}. For any subset $\J \subseteq [p]$, the cell count of $\bs x_{\J}$ is $\bs n_{\J}(\bs x_{\J}) = \sum_{i=1}^n \textbf{1}_{\bs x_{i \J} = \bs x_{\J}}$, that is, the number of samples $\lb \bs x_{i \mathcal J} = (x_{ij})_{j \in \J} \rb$ generated from the cell $\bs x_{\J}$. Key properties of the cell counts include: $\sum_{\bs x_{\mathcal J}} \bs n_{\J}(\bs x_{\J}) = n$; and for any $\J^\prime \subset \J$, $\bs n_{\J^\prime}(\bs x_{\J^\prime}) = \sum_{i=1}^n \bs n_{\J}(\bs x_{\J}) \textbf{1}_{\bs x_{i \J^\prime} = \bs x_{\mathcal J^\prime}}$. The discrete tensor $\bs n = (\bs n(\bs x))_{\bs x \in \chi}$ is called the \textit{contingency table}, and we can obtain contingency tables for any $\J \subset [p]$ with $\bs n_{\J} = (\bs n(\bs x_{\J}))_{\bs x_{\J} \in \chi_{\J}}$, with $\chi_{\J} = \lb (x_{j}) : j \in \J, x_j \in [k_j] \rb$.

DBNs focus on modeling cell counts for each node and its parent set \citep{scutari2021bayesian}. Let $\bs n_{\Pa(j),j}(\bs x_{\Pa(j)}, x_j)$ be the cell count of $(\bs x_{\Pa(j)}, x_j)$; for any root $j$, these are the marginal cell counts $\bs n_{\Pa(j),j}(\bs x_{\Pa(j)}, x_j) = \bs n_j(x_j)$. Given any $\bs x_{\Pa(j)}$, let $\bs n_{j \mid \Pa(j)}(\cdot \mid \bs x_{\Pa(j)}) = (\bs n_{\Pa(j),j}(\bs x_{\Pa(j)},x_j))_{x_j \in [k_j]}$ be the conditional cell counts for the $j$th node. 
Under \eqref{eq:Bayesian-network}, the multinomial likelihood of the global contingency table factorizes according to
\begin{equation} \label{eq:multinomial-likelihood}
   f(\bs n \mid \bs \pi, G) \propto \prod_{j=1}^p \prod_{\bs x_{\Pa(j)}, x_j} \bs \pi_{j \mid \Pa(j)}(x_j \mid \bs x_{\Pa(j)})^{\bs n_{\Pa(j),j}(\bs x_{\Pa(j)},x_j)},
\end{equation}
where proportionality is conveyed in terms of $(\bs \pi,G)$. Equation \eqref{eq:multinomial-likelihood} makes no further assumptions on the conditional PMF for each node, i.e. we do not assume an additive relationship between $\bs x_{\Pa(j)}$ and $x_j$. Conditionally on $G$, the maximum likelihood estimators (MLEs) for any cell with non-zero counts are
\begin{equation} \label{eq:MLEs}
    \widehat{ \bs \pi}_{j \mid \Pa(j)}(x_j \mid \bs x_{\Pa(j)}) = \frac{\bs n_{\Pa(j),j}(\bs x_{\Pa(j)},x_j)}{\bs n_{\Pa(j)}(\bs x_{\Pa(j)})},
\end{equation}
that is, the proportions of cell counts for the subgraph $H_j = (\Pa(j)\cup \lb j \rb, F_j) \subset G$, where $F_j = \lb (j^\prime,j): j^\prime \in \Pa(j) \rb$. These cell counts are important quantities for testing I-map conditions via chi-square tests \citep{agresti2002categorical}.

Estimating the conditional probabilities can be brittle to \textit{sparsity}, i.e. when $\bs n_{\Pa(j), j}(\bs x_{\Pa(j)},x_j)$ are small, motivating
a Bayesian approach. As \eqref{eq:multinomial-likelihood} is an exponential family, there exists a conjugate prior: assign independent Dirichlet distributions to $\bs \pi_{j \mid \Pa(j)}(\cdot \mid \bs x_{\Pa(j)})$ for all categories $\bs x_{\Pa(j)}$ \citep{lindley1964bayesian, dawid1993hyper, castelletti2021equivalence}. The posterior mean of $\bs \pi_{j \mid \Pa(j)}(x_j \mid \bs x_{\Pa(j)})$, conditional on $G$, is a shrinkage estimator, taking values between the cell proportions \eqref{eq:MLEs} and the prior mean. 
Hence, we can think of the Dirichlet hyperparameters as being ``prior cell counts" for variable $j$ and its parents. Beyond this interpretation, it is difficult to choose their values in practice, in part due to the vast dimension of the hyperparameters. The conjugate prior for \eqref{eq:multinomial-likelihood} introduces $\sum_{j=1}^p k_j \prod_{j^\prime \in \Pa(j)} k_{j^\prime}$ prior cell counts, one for each cell in $\bs n_{\Pa(j),j}$. Of course, one could simply set the prior cell counts to be equal across all the cells, reducing the problem to just one hyperparameter, but this choice can lead to brittleness.

\begin{figure}[t]
    \centering
    \includegraphics[scale=0.22]{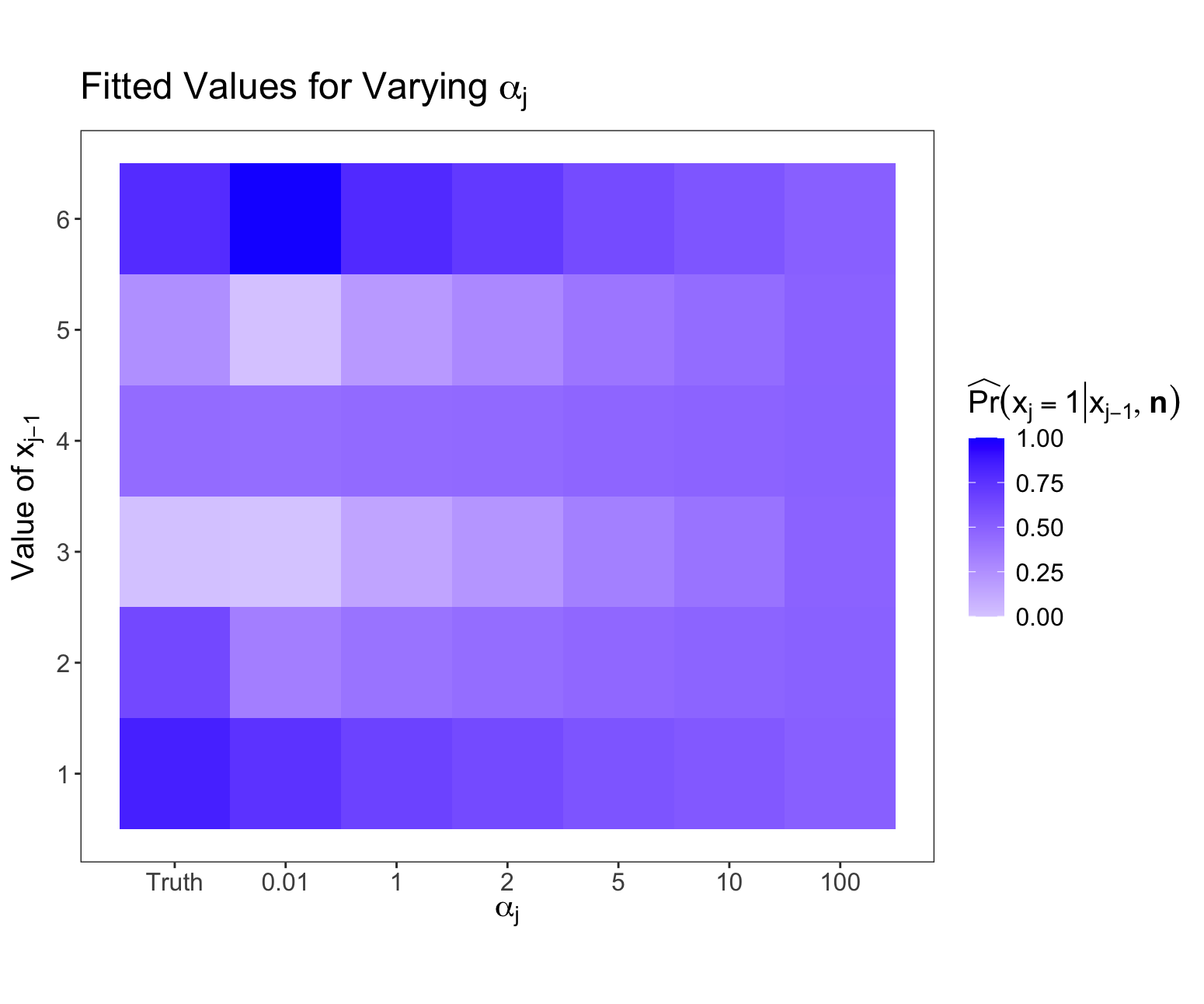}
    \caption{Fitted values $\widehat{\Pr}(x_j=1 \mid x_{j-1}, \bs n)$ for MLEs $\widehat{\bs \pi}_{j \mid j-1}(1) = (3/4, 1/3, 0/5, 3/7, 0/3, 3/3)$ and $\alpha_j$ varying in $\{ 0.01, 1, 2, 5, 10, 100 \}$, with comparison to the true values of $\Pr(x_j=1 \mid x_{j-1})$ in the leftmost column.}
    \label{fig:fitted-values-example}
\end{figure}

For instance, Figure \ref{fig:fitted-values-example} shows the posterior predictive probabilities of a binary variable $x_j$ with a single parent, $x_{j-1}$, which has $k_{j-1} = 6$ categories. We set $n=25$, and the MLEs for $\Pr(x_j=1 \mid x_{j-1})$ are $(3/4, 1/3, 0/5, 3/7, 0/3, 3/3)$. We assume that $(\bs \pi_{j \mid j-1}(0 \mid x_{j-1}), \bs \pi_{j \mid j-1}(1 \mid x_{j-1}))  \sim \tx{Dir}(\alpha_j, \alpha_j)$ independently, and we vary $\alpha_j \in \lb 0.01, 1, 2, 5, 10, 100 \rb$. For comparison, the first column of Figure \ref{fig:fitted-values-example} are the true simulated conditional probabilities, $\Pr(x_{j}=1 \mid x_{j-1})$. The sparse counts make $\alpha_j$ very influential on the fitted values. A small value of $\alpha_j$ leads to a better estimate of $\Pr(x_j=1 \mid x_{j-1}=6)$, but a worse estimate of $\Pr(x_j=1 \mid x_{j-1}=2)$. In addition, letting $\alpha_j$ grow very large results in trivial behavior, setting $\widehat{\Pr}(x_j=1 \mid x_{j-1}, \bs n) = 1/2$. 

We pursue an alternate strategy, retaining the nonparametric likelihood in \eqref{eq:multinomial-likelihood} but circumventing the high dimensionality and brittleness of the Dirichlet-multinomial. First, we relax the assumption that all of the conditional probabilities are statistically independent. Returning to the example in Figure \ref{fig:example-dag}, suppose that $k_j = 2$ for all $j=1, \dots, 5$. Since $\Pa(4) = \lb 1,3 \rb$, the conditional probabilities of $x_4$ can be visualized as
\begin{equation} \label{eq:conditional-probs-example}
    \bs \pi_{4 \mid \lb 1,3 \rb} = 
    \begin{pmatrix}
        \bs \pi_{4 \mid \lb 1,3 \rb }(0 \mid 0,0) & \bs \pi_{4 \mid \lb 1,3 \rb }(1 \mid 0,0) \\
        \bs \pi_{4 \mid \lb 1,3 \rb }(0 \mid 1,0) & \bs \pi_{4 \mid \lb 1,3 \rb }(1 \mid 1,0) \\
        \bs \pi_{4 \mid \lb 1,3 \rb }(0 \mid 0,1) & \bs \pi_{4 \mid \lb 1,3 \rb }(1 \mid 0,1) \\
        \bs \pi_{4 \mid \lb 1,3 \rb }(0 \mid 1,1) & \bs \pi_{4 \mid \lb 1,3 \rb }(1 \mid 1,1) \\
    \end{pmatrix}.
\end{equation}
The rows of $\bs \pi_{4 \mid \lb 1,3 \rb}$ are all probability vectors that describe the \textit{same events}, i.e., whether $x_4$ takes a value of $0$ or $1$, but across different possible values of the parents. Under the standard Dirichlet-multinomial model, the rows are assumed to be independent before observing the cell counts. Instead, it is more realistic to generate dependent conditional probabilities, since each row of $\pi_{4 \mid \{1,3\}}$ describes the same phenomena but conditioned on different events. A natural way to induce prior dependence is through a hierarchical model in which conditional probabilities are independent conditional on a common prior mean. In sparse regimes, the conditional probabilities for cells with small counts will shrink towards the common prior mean, combining information across the rows of $\bs \pi_{4 \mid \lb 1,3 \rb}$. The Dirichlet-multinomial model would shrink these probabilities towards row-specific concentrations that are entirely specified by the user, such as in Figure \ref{fig:fitted-values-example}.

Additionally, conditional on $G$, the parents of node $j$ partition the samples into $K_{\Pa(j)} = \prod_{j^\prime \in \Pa(j)} k_{j^\prime}$ groups, with the size of each group given by $\bs n_{\Pa(j)}(\bs x_{\Pa(j)})$ for all categories $\bs x_{\Pa(j)}$. The groups are defined by the allocation of observations to cells based on the values of $\bs x_j$ and $\bs x_{\Pa(j)}$. We demonstrate this by focusing on the transition kernel of the stochastic process of cell counts, defined by advancing from node-to-node on the DAG. 

We show that the transition kernel for $\bs n_j$ conditional on the cell counts of its parents is a Poisson-multinomial distribution \citep{chen1997statistical,lin2022poisson}, a generalization of the multinomial distribution that allows for \textit{distinct} event probabilities. For an experiment with $k$ outcomes, we denote the Poisson-multinomial distribution as $\tx{P-Mult}(n, P)$, where $n$ is the number of trials (as in the typical multinomial set-up), and $P$ is a $n \times k$ matrix whose rows are probability vectors. The Poisson-multinomial simplifies to the multinomial when the rows of $P$ are all equal.

\begin{prop} \label{prop:poisson-multinomial}
        Let $G$ be a DAG and $\lb x_i \rb$ a multinomial sample from a DBN defined by $G$. The conditional distribution of $\bs n_j$ given its parent contingency table $\bs n_{\Pa(j)}$ is
$            \bs n_j \mid \bs n_{\Pa(j)}, \bs \pi, G \sim \tx{P-Mult}(n, P_j),$
        where $P_j = (\bs p_{j1}, \dots, \bs p_{jn})^T$ and $\bs p_{ji} = \bs \pi_{j \mid \Pa(j)}(\cdot \mid \bs x_{i\Pa(j)})$.
\end{prop}

The rows of $P_j$ in Proposition \ref{prop:poisson-multinomial} have a specific form: they must be one of the rows of $\bs \pi_{j \mid \Pa(j)}$ by construction. In the context of Figure \ref{fig:example-dag}, this means that $\bs p_{ji}$ must be one of the rows of $\bs \pi_{4 \mid \lb 1,3 \rb}$ for all $i=1, \dots, n$. Hence, we expect certain rows of $P_j$ to be identical, resulting in clustering of the observations into $K_{\Pa(j)}$ groups, similar to Bayesian hierarchical models for grouped data \citep{gelman2013bayesian}. In light of this perspective, we specify an appropriate hierarchical framework for learning the conditional probabilities in DBNs.

\subsection{Hierarchical Model of Dirichlet Kernels}
Following the group structure we discussed in the previous section, we assume that the conditional probabilities of each node are drawn from hierarchical Dirichlet kernels:
\begin{equation} \label{eq:general-DAG-HiDDeN}
    \begin{gathered}
        \bs \pi_{j \mid \Pa(j)}(\cdot \mid \bs x_{\Pa(j)}) \overset{ind}{\sim} \tx{Dir}(\bs t_j); \tx{ for all } \bs x_{\Pa(j)} \in \chi_{\Pa(j)}; \; \:
        \bs t_j(x_j) \overset{ind}{\sim} \tx{Gamma}(\rho_j/k_j, b_j);
    \end{gathered}
\end{equation}
where $\bs t_j = (\bs t_j(1), \dots, \bs t_j(k_j)) \in \mathbb{R}^{k_j}$ and $\rho_j,b_j>0$. We view $\bs t_j$ as an unnormalized latent prior mean, which defines a common prior expectation for all $\bs \pi_{j \mid \Pa(j)}(\cdot \mid \bs x_{\Pa(j)})$. If $j$ is a root, we specify \eqref{eq:general-DAG-HiDDeN} as a hierarchical prior for $\bs \pi_j(\cdot)$. We highlight that \eqref{eq:general-DAG-HiDDeN} can be applied to any DAG and for any number of categories in the different categorical variables. We refer to the prior for $\bs \pi$ in \eqref{eq:general-DAG-HiDDeN} as a Hierarchical Directed Dirichlet Network (HiDDeN). 

Let $\beta_j = \sum_{j=1}^{k_j} \bs t_j(x_j)$ and $\btildepi_j = \bs t_j/\beta_j$. Returning to our running example in Figure \ref{fig:example-dag}, HiDDeN will introduce prior means for each non-root: $\btildepi_2$, $\btildepi_3$, $\btildepi_4$, and $\btildepi_5$. A priori, the rows of $\bs \pi_{4 \mid \lb 1,3 \rb}$ in \eqref{eq:conditional-probs-example} are centered on $\btildepi_4$, meaning $\E[\bs \pi_{4 \mid \lb 1,3 \rb} \mid  \btildepi_4] = (\btildepi_4, \btildepi_4, \btildepi_4, \btildepi_4)^T$. One can show that $\beta_j$ controls variation in the conditional probabilities from their common prior means; as $\beta_j \to \infty$, then $\pi_{j \mid \Pa(j)}(\cdot \mid \bs x_{\Pa(j)}) \to \btildepi_j$. Furthermore, our representation in \eqref{eq:general-DAG-HiDDeN} is equivalent to specifying Dirichlet and gamma priors for $\btildepi_j$ and $\beta_j$, respectively \citep{das2024blocked}. For roots, \eqref{eq:general-DAG-HiDDeN} can be interpreted as a prior distribution for a concentration parameter in the typical Dirichlet prior. While we focus on non-roots in this and the following sections, introducing latent variables $\bs t_j$ for all nodes is relevant to our approach to structure learning in Section \ref{section:structure-learning}.

Let $\xnew_j$ be a new observation from the $j$th node \textit{after} observing the data. Then, the posterior predictive probability of $\xnew_j$ is
\begin{equation} 
\begin{gathered} \label{eq:predictive-prob}
    \Pr(\bs x_j^{\tx{new}} = x_j \mid \bs x_{\Pa(j)}, \bs n,G) = \E \bigg [ \frac{\bs t_j(x_j) + \bs n_{\Pa(j),j}(\bs x_{\Pa(j)}, x_j)}{\beta_j + \bs n_{\Pa(j)}(\bs x_{\Pa(j)})} \bigg | \bs n_{\Pa(j),j},G \bigg ].
\end{gathered}
\end{equation}
For small values of $\bs n_{\Pa(j),j}(\bs x_{\Pa(j)},x_j)$, \eqref{eq:predictive-prob} shrinks toward $\E[\bs t_j(x_j)/\{\beta_j + \bs n_{\Pa(j)}(\bs x_{\Pa(j)}\}) \mid \bs n_{\Pa(j),j}, G]$, concatenating information on all conditional probabilities $ \{\bs \pi_{j \mid \Pa(j)}(x_j \mid \bs x_{\Pa(j)}) \}_{\bs x_{\Pa(j)}}$. In the context of the example conditional probabilities in \eqref{eq:conditional-probs-example}, HiDDeN learns common structure in the rows of $\bs \pi_{4 \mid 1,3}$ via $\bs t_4$, which is then used to infer sparse cells and avoid hyperparameter sensitivity. We explicitly show how information is shared in the following section. 

\section{MALA-Within-Gibbs Sampler} \label{section:posterior}
Standard MCMC samplers for finite approximations to the HDP, such as that of \cite{teh2006sharing}, could be applied to HiDDeN by reparameterizing \eqref{eq:general-DAG-HiDDeN}. However, these algorithms are not motivated by the setting of DBNs and assume that the cell counts are derived from latent variables. Instead, we propose a novel marginal Metropolis-within-Gibbs sampler that uses the setting of DBNs to ensure efficiency. Importantly, our approach does not rely on variational inference, which has previously been applied to hierarchical Dirichlet kernels on DAGs \citep{azzimonti2019hierarchical,azzimonti2022bayesian}.

As we do not need $\bs \pi_{j \mid \Pa(j)}(\cdot \mid \bs x_{\Pa(j)})$ to compute the fitted values \eqref{eq:predictive-prob}, we marginalize these variables out from the likelihood, which in turn reduces the size of the parameter space. After marginalizing over these quantities, the likelihood contribution of the nodes are $f(\bs n \mid \bs t, G) \propto \prod_{j=1}^p f(\bs n_{j \mid \Pa(j)} \mid \bs n_{\Pa(j)}, \bs t_j)$, where
\begin{equation}
    \begin{gathered} \label{eq:n-given-pi-tilde}
    f(\bs n_{j \mid \Pa(j)} \mid \bs n_{\Pa(j)}, \bs t_j) = \prod_{\bs x_{\Pa(j)} \in \chi_{\Pa(j)}} \frac{\Gamma(\beta_j)}{\Gamma(\bs n_{\Pa(j)}(\bs x_{\Pa(j)}) + \beta_j)} \prod_{x_j=1}^{k_j} \frac{\Gamma(\bs t_j(x_j) +  \bs n_{\Pa(j),j}(\bs x_{\Pa(j)}, x_j)}{\Gamma(\bs t_j(x_j))}.
    \end{gathered}
\end{equation}
If $\Pa(j) = \emptyset$, we introduce a dummy variable for the parents of $j$ so that $\chi_{\Pa(j)} = \lb 1 \rb$. Hence, our algorithm is applicable to both roots and non-roots. By construction of our prior, the MCMC updates of each $\bs t_j$ can be carried out independently. The full joint distribution for a single node is obtained by multiplying the prior distribution for $\bs t_j$ in \eqref{eq:general-DAG-HiDDeN} to its associated term in \eqref{eq:n-given-pi-tilde}:
\begin{equation}
    \begin{gathered} \label{eq:das-full-joint}
        f(\bs n_{j \mid \Pa(j)}, \bs t_j \mid \bs n_{\Pa(j)}, G) \propto \prod_{\bs x_{\Pa(j)}} \frac{\Gamma\left(\beta_j \right)}{\Gamma(\bs n_{\Pa(j)}(\bs x_{\Pa(j)}) + \beta_j )} \\ 
    \times \prod_{x_j=1}^{k_j} \frac{ \exp \lb-b_j \bs t_j(x_j) \rb \bs t_j(x_j)^{\rho_j/k_j-1}}{\Gamma(\bs t_j(x_j))^{K_{\Pa(j)}}} \lb \prod_{\bs x_{\Pa(j)}} \Gamma(\bs n_{\Pa(j),j}(\bs x_{\Pa(j)}, x_j) + \bs t_j(x_j)) \rb,
    \end{gathered}
\end{equation}
where $K_{\Pa(j)} = |\chi_{\Pa(j)}| = \prod_{j^\prime \in \Pa(j)} k_{j^\prime}$. The distribution in \eqref{eq:das-full-joint} is split into two terms. The first is a function of the sum \textit{across} categories in the $j$th node, $\beta_j = \sum_{x_j=1}^{k_j}\bs t_j(x_j)$. The second term factorizes as a product over these categories. Using elementary properties of the Beta function, we obtain that the first term can be decomposed as 
\begin{gather*}
    \frac{\Gamma \left( \beta_j \right)}{\Gamma(\bs n_{\Pa(j)}(\bs x_{\Pa(j)}) + \beta_j)}
    = \frac{\int_{0}^{1} \bs u_j(\bs x_{\Pa(j)})^{\beta_j - 1} (1-\bs u_j(\bs x_{\Pa(j)}))^{\bs n_{\Pa(j)}(\bs x_{\Pa(j)}) - 1} d\bs u_j(\bs x_{\Pa(j)})}{\Gamma(\bs n_{\Pa(j)}(\bs x_{\Pa(j)}))}
\end{gather*}
for all $\bs x_{\Pa(j)}$. This decomposition suggests the Gibbs sampler given in Algorithm \ref{alg:prior-means} to sample $\bs t_j$.

\begin{algorithm}[t]
\caption{MCMC Update of Concentrations}
\label{alg:prior-means}
\begin{algorithmic}[1]
\Require{Parent set $\Pa(j) \subset [p] \setminus \lb j \rb$, current concentrations $\bs t_j$, hyperparameters $b_j,\rho_j>0$.}
    \For{$\bs x_{\Pa(j)} \in \chi_{\Pa(j)}$}
        \State Sample $\bs u_j(\bs x_{\Pa(j)}) \sim \tx{Beta}(\beta_j, \bs n_{\Pa(j)}(\bs x_{\Pa(j)}))$, where $\beta_j = \sum_{x_j} \bs t_j(x_j)$;
    \EndFor
    \For{$x_j \in [k_j]$} 
        \State Sample $\bs t_j(x_j) \sim h_{j,x_j}$, where
        \begin{equation}  \label{eq:h_j}
        h_{j,x_j}(t) \propto \frac{ \exp \lb-b_j t \rb t^{\rho_j/k_j-1}}{\Gamma(t)^{K_{\Pa(j)}}} \lb  \prod_{\bs x_{\Pa(j)}} \Gamma(\bs n_{\Pa(j),j}(\bs x_{\Pa(j)}, x_j) + t) \bs u_j(\bs x_{\Pa(j)})^{t} \rb
        \end{equation}
    \EndFor
\State \Return Updated values of $\bs t_j$.
\end{algorithmic}
\end{algorithm}

Each step in the Gibbs sampler can be done using parallel computing. The auxiliary variables $\bs u_j(\bs x_{\Pa(j)})$ are independent across the values of $\bs x_{\Pa(j)}$; the entries of the latent concentrations $\bs t_j$ are independent across the values of $x_j$. Furthermore, information is shared across the parent set by the product in \eqref{eq:h_j}: the cell counts $\{\bs n_{\Pa(j),j}(\bs x_{\Pa(j)},x_j)\}_{\forall \: \bs x_{\Pa(j)}}$ contribute to the expression of $h_{j,x_j}$.
We show that $h_{j,x_j}(t)$ is log-concave under mild conditions on marginal cell counts and hyperparameters. 
\begin{theorem} \label{thm:log-concave}
    Let $\Pa(j) \subseteq [p] \setminus \{ j \}$. The function $h_{j,x_j}(t)$ is log-concave in $t$ for all values of the auxiliary variables $\{ \bs u_j(\bs x_{\Pa(j)}) \}_{\bs x_{\Pa(j)} \in \chi_{\Pa(j)}}$ if $\rho_j \geq k_j$ and $\bs n_j(x_j) >0$.
\end{theorem}
Log-concavity is shown by taking derivatives of the logarithm of \eqref{eq:h_j}. The first two derivatives are
\begin{gather}
    \frac{d \log h_{j,x_j}(t)}{dt} = -b_j + \frac{\rho_j/k_j-1}{t} + \sum_{\bs x_{\Pa(j)}} \{ \psi(\bs n_{\Pa(j),j}(\bs x_{\Pa(j)}, x_j) + t) - \psi(t) + \log \bs u_j(\bs x_{\Pa(j)}) \} \label{eq:t-gradient} \\
    \frac{d^2 \log h_{j,x_j}(t)}{dt^2} = \frac{1 - \rho_j/k_j}{t^2} + \sum_{\bs x_{\Pa(j)}} \{ \psi^{(1)}(\bs n_{\Pa(j),j}(\bs x_{\Pa(j)}, x_j) + t) - \psi^{(t)}(t) \}, \label{eq:t-hessian}
\end{gather}
where $\psi(t) = \frac{d}{dt}\log \Gamma(t)$ and $\psi^{(1)}(t) = \frac{d^2}{dt^2}\log \Gamma(t)$ are the digamma and trigamma functions, respectively. The condition on $\rho_j$ is needed to make the first term in \eqref{eq:t-hessian} negative, while $n_j(x_j)>0$ is a sufficient condition for the second term in \eqref{eq:t-hessian} to be negative since the trigamma function is decreasing. These conditions are minimal are practice. We typically set $\rho_j$ to be some value larger than $k_j$, such as $\rho_j = k_j+1$, to satisfy Theorem \ref{thm:log-concave}. The second condition states that we require that category $x_j$ is represented by at least one observation, which is trivial to verify.

In light of Theorem \ref{thm:log-concave}, we propose using the Metropolis-adjusted Langevin algorithm (MALA) to simulate from \eqref{eq:h_j}, leveraging the appealing properties of MALA for sampling log-concave densities \citep{dwivedi2019log}. The Supplemental Material explains in detail the MALA implementation for the HiDDeN model. Given a current value $\bs t_j^{(r)}(x_j)$, the Langevin proposal generates $\bs t_j^*(x_j)$ with
\begin{equation} \label{eq:langevin-proposal}
    \bs t_j^*(x_j) = \bs t_j^{(r)}(x_j) + \frac{\bs \epsilon_j(x_j)^2}{2}  \frac{d \log h_{j,x_j}(\bs t_j^{(r)}(x_j))}{dt} + \bs \epsilon_j(x_j) z^{(r)},
\end{equation}
where $\bs \epsilon_j(x_j)>0$ is the step size and $z^{(r)} \sim \N(0,1)$. MALA pools information across the parent categories using the derivative \eqref{eq:t-gradient}: all of the node-parent cell counts $\bs n_{\Pa(j),j}(\bs x_{\Pa(j)},x_j)$ are added to $\bs t_j(x_j)^{(r)}$ and summed over. In addition, we also achieve pooling of information across the values of node $j$ with $\bs u_j(x_{\Pa(j)})$. The algorithm requires the specification of step size parameters $\bs \epsilon_j(x_j)>0$ for all $x_j \in [k_j]$. The step size cannot be naively chosen, e.g. setting $\bs \epsilon_j(x_j)$ too large can result in negative Langevin proposals, which are automatically rejected in the Metropolis step. We adaptively adjust the step size using the acceptance probabilities for $\bs t(x_j)$ and MCMC diagnostics for the log-posterior, favoring a balance between successful proposals and a sufficient exploration of the parameter space.

The complexity of Algorithm \ref{alg:prior-means} depends on multiple factors. The contingency tables can be derived in advance and stored as $\bs n_{\Pa(j)}$ (parent cell counts) and $\bs n_{\Pa(j),j}$ (parent-child cell counts). The computation time for $\bs n_{\Pa(j),j}$ is bounded above by $\mathcal O(nK_{\Pa(j)}k_j)$, and $\bs n_{\Pa(j)}$ is obtained by summing the entries in $\bs n_{\Pa(j),j}$, which is $\mathcal O(K_{\Pa(j)}k_j)$. 
The number of categories in the parents, $K_{\Pa(j)}$, determines the number of auxiliary variables to sample and the computation time for the gradient in \eqref{eq:t-gradient}. $K_{\Pa(j)}$ is influenced both by the cardinality of the parent set and by the number of categories in each parent. For example, $K_{\Pa(j)} = 16$ when node $j$ has $4$ parents with $4$ categories each, and when node $j$ has $2$ parents with $2$ categories and $8$ categories, respectively. Similarly, the number of categories in $j$, or $k_j$, influences the calculation of $\beta_j$ and the sampling of $\bs t_j$. Therefore, the complexity of doing one sweep of Algorithm \ref{alg:prior-means} is $\mathcal O(K_{\Pa(j)}k_j)$, assuming the tables are made before running MCMC.

\section{Structure Learning} \label{section:structure-learning}

\subsection{Edge Indicators}
We first address the problem of deriving an unknown parent set of a single node $j$ with little prior information. For nodes $j$ and $j^\prime$, let $z_{j^\prime j} = \textbf{1}_{j^\prime \in \Pa(j)}$ be the indicator of $(j^\prime,j) \in E$. In addition, suppose that we have access to a set of candidate parents $\Ca(j) \subset [p]\setminus \lb j \rb$, so that $f\lb \Pa(j) \subseteq \Ca(j) \rb = 1$, meaning that $\Pa(j) = \lb j^\prime \in \Ca(j) : z_{j^\prime j} = 1 \rb$. Conditional on $\Ca(j)$, we model $z_{j^\prime j} \sim \tx{Ber}(\gamma_j)$, where $0 < \gamma_j < 1$ controls the size of the parent set, or \textit{in-degree}, of node $j$, since $|\Pa(j)| = \sum_{j^\prime \in \Ca(j)} z_{j^\prime j} \sim \tx{Bin}(|\Ca(j)|, \gamma_j)$. We collect the indicators into a vector $\bs z_j = (z_{j^\prime})_{j^\prime \in \Ca(j)}$. To complete the specification, we also assume $\gamma_j \sim \tx{Beta}(c_j, d_j)$ for $c_j,d_j>0$. Importantly, our formulation allocates non-zero prior probability to all subsets of $\Ca(j)$.

The MCMC update for $z_{j^\prime j}$ is a sample from a Bernoulli distribution, with probabilities
\begin{equation} \label{eq:z-update}
\begin{gathered} 
    f(z_{j^\prime j} = 0 \mid -) \propto \bigg \{ \sum_{j^{\prime \prime} \in \Ca(j)^{-j^\prime}}(1- z_{j^{\prime \prime}j}) + d_j \bigg \} f(\bs n_{j \mid \Pa(j)^{-j^\prime}} \mid \bs n_{\Pa(j)^{-j^\prime}}, \bs t_j) \\
    f(z_{j^\prime j} = 1 \mid -) \propto  \bigg \{ \sum_{j^{\prime \prime} \in \Ca(j)^{-j^\prime}} z_{j^{\prime \prime}j} + c_j \bigg \} f(\bs n_{j \mid \Pa(j)} \mid \bs n_{\Pa(j)}, \bs t_j);
\end{gathered}
\end{equation}
where $\Ca(j)^{-j^\prime} = \Ca(j) \setminus \lb j^\prime \rb$, $z_j^{-j^\prime} = (z_{j^{\prime \prime} j})_{j^{\prime \prime} \in \Ca(j)^{-j^\prime}}$, $\Pa(j)^{-j^\prime} = \{ j^{\prime \prime } \in \Ca(j) \setminus \lb j^\prime \rb : z_{j^{\prime \prime}j}=1 \}$, and $\Pa(j) = \Pa(j)^{-j} \cup \lb j \rb $. By convention, if $\Pa(j)^{-j} = \emptyset$, we set $\bs \chi_{\Pa(j)^{-j^\prime}} = \lb 1 \rb$ and $\bs n_{\Pa(j)^{-j^\prime}}(1)=n$. With $\gamma_j$ marginalized out in \eqref{eq:z-update}, the hyperparameters $c_j$ and $d_j$ influence the distribution of the in-degree of $j$; for example,  
if $c_j/d_j \to \infty$ then $z_{j^\prime j} \to 1$. Sampling from \eqref{eq:z-update} for all $j\prime \in \Ca(j)$ adds or removes edges from $\Pa(j)$. 

A single MCMC update of $\bs z_j$ 
is detailed in Algorithm \ref{alg:edge-indicators}. Once each entry of $\bs z_j$ is updated, we proceed according to the Gibbs sampler in Algorithm \ref{alg:prior-means} with the updated parent set $\Pa(j) = \lb j^\prime \in \Ca(j) : z_{j^\prime j} = 1  \rb$. Each iteration has complexity bounded by $\mathcal O( n K_{\Pa(j)} k_j + k_j + K_{\Pa(j)} + |\Ca(j)-1| )$, assuming that the contingency tables are recalculated during every loop through $\Ca(j)$. If the previous $z_{j^\prime j}=1$, updating that edge indicator is more efficient since $\Pa(j)^{-j}$ is obtained by summing the entries in the contingency table. 

\begin{algorithm}[t]
\caption{MCMC Update of Edge Indicators}
\label{alg:edge-indicators}
\begin{algorithmic}[1]
\Require{Candidate set $\Ca(j) \subset [p] \setminus \lb j \rb$, current indicators $\bs z_j$, latent concentrations $\bs t_j$, and hyperparameters $c_j,d_j>0$.}
    \For{$j^\prime \in \Ca(j)$}
        \State $\Pa(j)^{-j^\prime} = \{ j^{\prime \prime} \in \Ca(j)^{-j} : z_{j^{\prime \prime} j}=1 \}$ and $\Pa(j) = \Pa(j)^{-j^\prime} \cup \{ j^\prime \}$;
        \State Sample $z_{j^\prime j} \sim \tx{Ber}(\rho_{j^\prime j})$, where $\rho_{j^\prime j}=f(z_{j^\prime j}=1 \mid -)$ from \eqref{eq:z-update};
    \EndFor
\State \Return $\bs z_j$.
\end{algorithmic}
\end{algorithm}

\subsection{Bayesian Parent Selection} \label{sect:local-approach}
In contrast to the setting of Algorithm \ref{alg:edge-indicators}, suppose that we prespecify a moderate number of candidate parents $\Pa(j)_1, \dots, \Pa(j)_{M_j}$ and prior probabilities $\Pr(\Pa(j)_1), \dots, \Pr(\Pa(j)_{M_j})$, which can be chosen to favor certain nodes as parents or restrict the degree of $j$. For example, in medicine, $\{\Pa(j)_m\}_{m=1}^{M_j}$ can represent different combinations of interventions for a disease, while node $j$ is an outcome. Default prior probabilities can be chosen as $\Pr(\Pa(j)_m) = 1/M_j$ for all $m \in [M_j]$. Unlike using edge indicators, this setup allows for direct modeling of the parent set and can lead to a markedly smaller space of network structures for the MCMC algorithm to explore over. Also, under this model we update the entire parent set in a single step rather than passing through all possible edges.

We generate draws from the posterior $\Pr(\Pa(j) \mid \bs n)$ by iteratively updating $\Pa(j)$ within the set of candidate parents. 
We compute the conditional probability of the $m$th candidate parent as $\Pr(\Pa(j)_m \mid - ) \propto \Pr(\Pa(j)_m) f(\bs n_{j \mid \Pa(j)_m} \mid \bs n_{\Pa(j)_m}, \bs t_j).$
We then set $\Pa(j)=\Pa(j)_m$ with probability $\Pr(\Pa(j)_m \mid - )$. Taking the logarithm of $\Pr(\Pa(j)_m \mid - )$ returns the Bayesian Dirichlet (BDE) graph score \citep{heckerman1995learning,friedman1998bayesian} of $\Pa(j)_m$ with the latent concentrations in place of the prior sample sizes. Algorithm \ref{alg:BPS} shows the update of $\Pa(j)$ in our MCMC sampler, which we then use to update $\bs t_j$ in Algorithm \ref{alg:prior-means}. In contrast to Algorithm \ref{alg:edge-indicators}, we can store the parent and parent-child contingency tables $\bs n_{\Pa(j)_m}$ and $\bs n_{\Pa(j)_m,j}$ beforehand, increasing efficiency. This makes the complexity of one pass through Algorithm \ref{alg:BPS} $\mathcal O(M_j k_j K_{\Pa(j)} )$.

\begin{algorithm}[t]
\caption{MCMC Update for Bayesian Parent Selection}
\label{alg:BPS}
\begin{algorithmic}[1]
\Require{Candidate parent sets $\lb \Pa(j)_m \rb_{m=1}^{M_j}$, prior probabilities $\lb \Pr(\Pa(j)_m) \rb_{m=1}^{M_j}$, and latent concentrations $\bs t_j$.}
    \State Sample $m_j \gets  \tx{Cat}(M_j, (\Pr(\Pa(j)_m | -) )_{m=1}^{M_j}) $, where the parent probabilities are $\Pr(\Pa(j)_m | - ) \propto \Pr(\Pa(j)_m) f(\bs n_{j \mid \Pa(j)_m} | \bs n_{\Pa(j)_m}, \bs t_j)$;
    \State $\Pa(j) \gets \Pa(j)_{m_j}$
\State \Return $\Pa(j)$.
\end{algorithmic}
\end{algorithm}

\subsection{MCMC Estimators for Graphical Quantities} \label{sect:mcmc-graph-estimators}

Alternating Algorithms \ref{alg:prior-means} and \ref{alg:edge-indicators} or \ref{alg:BPS} for $R>0$ iterations returns posterior draws of the parent sets, denoted $\Pa(j)^{(r)}$ for $r=1, \dots, R$; the binary edge indicators $\bs z_j^{(r)}$, where $z_{j^\prime j}^{(r)} = 1$ if and only if $j^\prime \in \Pa(j)^{(r)}$; and the latent concentrations $\bs t_j^{(r)}$. Importantly, the conditions of Theorem \ref{thm:log-concave} depend only on the marginal cell counts and hyperparameters, not the makeup of the parent set. Therefore, we retain efficient sampling of $\bs t_j$ even when the parent set is treated as unknown.

Posterior edge probabilities can be used to represent posterior uncertainty in graph learning; for example, using line widths that increase with edge probabilities. In addition, there is often particular interest in the directed relationships between specific variables. For example, in the causal inference setting, one may want to infer the weight of evidence of a directed edge between an exposure or treatment and an outcome variable.
The posterior probabilities are estimated using the binary edge indicators, $\Pr(z_{j^\prime j} = 1 \mid \bs n) \approx \widehat{\Pr}(z_{j^\prime j} = 1 \mid \bs n) =  (1/R)\sum_{r=1}^{R}z_{j^\prime j}^{(r)}$. We approximate the posterior probability of an entire parent set $\Pa(j)$ with $\Pr(\Pa(j) \mid \bs n) \approx \widehat \Pr(\Pa(j) \mid \bs n) = \frac{1}{R} \sum_{r=1}^{R} \textbf{1}_{\Pa(j)^{(r)} = \Pa(j)}$. Under appropriate loss functions, estimators $\widehat{\Pa}(j)$ can be calculated from $\widehat{\Pr}(\Pa(j) \mid \bs n)$. Reasonable estimators include the maximum a posteriori (MAP) estimator, which chooses the combination of parents having the highest posterior probability, or the median probability model \citep{barbieri2004optimal}, which chooses all parents having $\widehat \Pr(z_{j^\prime j} = 1 \mid \bs n) > 0.5$.

Finally, we obtain the predictive PMF of $\xnew$ by taking the expectation of \eqref{eq:predictive-prob} with respect to the posterior distribution of $\Pa(j)$:
\begin{equation}
    \begin{gathered} \label{eq:predictive-prob-unknown-DAG}
        \Pr(\xnew_j = x_j \mid \bs x_{\Ca(j)}, \bs n) \approx \frac{1}{R} \sum_{r=1}^{R} \frac{\bs t_j^{(r)}(x_j) + \bs n_{\Pa(j)^{(r)},j}(\bs x_{\Pa(j)^{(r)}}, x_j)}{\beta_j^{(r)} + \bs n_{\Pa(j)^{(r)}}(\bs x_{\Pa(j)^{(r)}})}.
    \end{gathered}
 \end{equation}
Here, we adopt the notation $\Pr(\xnew_j = x_j \mid \bs x_{\Ca(j)}, \bs n)$ to reflect that $\Pa(j)$ is unknown but constricted to be contained in a candidate set $\Ca(j)$. $\Ca(j)$ is prespecified by the user when applying Algorithm \ref{alg:edge-indicators}, and we can define $\Ca(j)$ from Algorithm \ref{alg:BPS} via $\Ca(j) = \{ j^\prime : \exists m \in [M_j] \tx{ with } j^\prime \in \Pa(j)_m \}$. \eqref{eq:predictive-prob-unknown-DAG} can be used to create a classifier for node $j$ that takes into account uncertainty in the conditional dependencies. Equation \eqref{eq:predictive-prob-unknown-DAG} shows that parent sets with sparse $(\bs x_{\Pa(j)}, \bs x_j)$ cells shrink the predictive probabilities towards $\bs t_j$. This may occur if, for instance, the posterior samples of $\Pa(j)$ are dense (contain a large number of parents), so the property of $\bs t_j(x_j)$ concatenating information across parent categories is advantageous in this setting.

\subsection{Inferring Markov Blankets and DAGs}

The algorithms and MCMC estimators detailed in the previous subsections can be generalized to infer Markov blankets and the overall graph $G$. The extension to Markov blankets is straightforward: let $\Ca(j) = [p] \setminus \{ j \}$, then run Algorithm \ref{alg:edge-indicators} or \ref{alg:BPS} (depending on the availability of prior information and the problem setting) to produce MCMC samples of the Markov blanket, denoted $\tx{Mb}(j)$, which we display as an undirected graph; see Section \ref{sect:Markov-blanket-sims} for an example. This process is then repeated for each node, resulting in posterior probabilities for $p$ undirected graphs defined on $p-1$ nodes.

To infer a DAG, we can use a similar strategy: conditional on the rest of the graph $G$, we define a set of candidate parents $\Ca(j)$ for a particular node $j$, update the associated parent set using Algorithms \ref{alg:edge-indicators} or \ref{alg:BPS}, and repeat these steps for each of the different nodes within the Gibbs sampler. In applying this approach, it is crucial to rule out cycles in defining the possible parents of node $j$ taking into account the current graph structure in other nodes. That is, if adding the edge $(j^\prime,j)$ to $E$ results in a cycle, we automatically set $z_{j^\prime j} = 0$ in Algorithm \ref{alg:edge-indicators} and $\max_{m:j^\prime \in \Pa(j)_m}\Pr(\Pa(j)_m \mid -) = 0$ in Algorithm \ref{alg:BPS}. In the special case in which there is a fixed stochastic ordering of the variables, we can apply HiDDeN in parallel across the nodes with $\Ca(j)=[j-1]$.

Alternatively, if we can narrow down consideration to a small-to-moderate number of valid DAGs $G_1,\ldots,G_M$, we can calculate posterior probabilities for these DAGs updating prior probabilities $\Pr(G_1), \dots, \Pr(G_M)$.
The posterior probabilities of each $G_m$ provide a model-based alternative to other graph scores such as BDE. We can update $G$ in a Gibbs sampler by computing $\Pr(G_m \mid - ) \propto \Pr(G_m)
    \prod_{j=1}^p f(\bs n_{j \mid \Pa(j)_m} \mid \bs n_{\Pa(j)_m}, \bs t_j)$
for all $m \in [M_j]$, then sampling a new DAG from these probabilities. The logarithm of $\Pr(G_m \mid - )$ is proportional to the BDE score of $G_m$ with $\{ \bs t_j \}_{j=1}^p$ acting as hyperparameters.

A single iteration of the above MCMC update is detailed in Algorithm \ref{alg:unknown-DAG}. The computation time of using Algorithm \ref{alg:unknown-DAG} for DAG inferences depends on the candidate DAGs $G_1, \dots, G_M$. For example, if all the parent sets remain constant in the $M$ candidates except for the parent set of a fixed $j$, then the likelihood contribution of every other node cancels out in the full conditional probability of $G_m$, and we are basically using Algorithm \ref{alg:BPS}. More generally, we only need to produce posterior samples of $\bs t_j$ for nodes whose parent sets change between the candidates. Similarly to Algorithm \ref{alg:BPS}, the contingency tables for each graph can be derived beforehand, improving computational efficiency, leading to the complexity of $\mathcal O( \sum_{m=1}^{M} \sum_{j=1}^{p} k_j K_{\Pa(j)_m} )$.

\begin{algorithm}[t]
\caption{MCMC Update for an Unkown DAG}
\label{alg:unknown-DAG}
\begin{algorithmic}[1]
\Require{Candidate DAGs $\lb G_m \rb_{m=1}^{M}$, prior probabilities $\lb \Pr(G_m) \rb_{m=1}^{M}$, and latent concentrations $\{\bs t_j\}_{j=1}^{p}$.}
    \State Sample $m \gets \tx{Cat}(M, (\Pr(G_m | -) )_{m=1}^{M_j}) $, $\Pr(G_m | - ) \propto \Pr(G_m)
    \prod_{j=1}^p f(\bs n_{j \mid \Pa(j)_m} | \bs n_{\Pa(j)_m}, \bs t_j)$;
    \State $G \gets G_{m}$
\State \Return $G$.
\end{algorithmic}
\end{algorithm}

As in the case of inferring a single parent set, Theorem \ref{thm:log-concave} implies log-concavity for all nodes $j$ in Algorithm \ref{alg:unknown-DAG}. The resulting MCMC samples of $G$ can be used to compute posterior DAG probabilities $\Pr(G_m \mid \bs n)$, as well as parental set probabilities, edge probabilities, and predictive PMFs using the same estimators in Section \ref{sect:mcmc-graph-estimators}. 

\section{Experiments} \label{section:simulations}

\subsection{Parameter Learning for Sparse Counts}

First, we explore how HiDDeN performs in inferring the conditional probabilities of a single binary node $j$ given its parent set. We restrict our attention to the case in which $|\Pa(j)|=1$ but vary the number of categories $K_{\Pa(j)}$, allowing us to explore the performance of HiDDeN for different degrees of sparsity. The parent cell counts are vectorized and simulated as $\bs n_{\Pa(j)} \sim \tx{Mult}(n, (1/K_{\Pa(j)}, \dots, 1/K_{\Pa(j)}))$, and we then generate the child cell counts conditional on the parents via $\bs n_{j \mid \Pa(j)}( 1 \mid x_{\Pa(j)]}) \sim \tx{Bin}(\bs n_{\Pa(j)}(x_{\Pa(j)}), q_{ x_{\Pa(j)}})$, where $q_{x_{\Pa(j)}}=2/3$ if $x_{\Pa(j)} \equiv 0 \: (\tx{mod } 2)$ and $q_{x_{\Pa(j)}}=1/3$ otherwise. In order to gauge different regimes of sample size and sparsity for a single node, we vary $K_{\Pa(j)} \in \lb 2,3,5,10 \rb$ and fix $n=100$. Our main goal is to estimate $\bs q = (q_{x_{\Pa(j)}})_{x_{\Pa(j)} \in [K_{\Pa(j)}]}$. For each value of $K_{\Pa(j)}$, we simulate $50$ independent datasets.

We compare to logistic regression of variable $j$ conditioned on its parent using maximum likelihood estimation (logit MLE), Bayesian methods (logit Bayes), a generalized additive model (logit GAM), as well as the MLEs for the conditional probabilities $\bs q$ \eqref{eq:MLEs}, and the predictive PMF for the Dirichlet-multinomial (DM) model. A HiDDeN point estimate of the conditional probabilities is obtained using the fitted values in \eqref{eq:predictive-prob-unknown-DAG}. The stepsize for HiDDeN is fixed at $\bs \epsilon_j = (0.5,0.5)$ for all simulations,
and the hyperparameters are fixed at $b_j=1$ and $\rho_j=2$, the latter chosen to satisfy log-concavity. 
Hyperparameters for all competitors are set to their defaults, and the Bayesian methods are run for $10,000$ MCMC samples. The averages and standard deviations for the RMSEs to $\bs q$ are displayed in Table \ref{tab:simulation-results}. When $K_{\Pa(j)} = 2$, the least sparse level, all methods perform similarly. For $K_{\Pa(j)}>2$, HiDDeN always returns the lowest average RMSE to $\bs q$, and the difference in average RMSE between HiDDeN and competitors steadily increases with $K_{\Pa(j)}$.

\begin{table}[t]
        \centering
        \begin{tabular}{ccccccc}
        \toprule
            $K_{\Pa(j)}$ & HiDDeN & Logit MLE  & Logit Bayes  & Logit GAM  & MLE  & DM \\
        \midrule
            $2$ & .06 (.034) & .06 (.034) & .06 (.034) & .06 (.034) & .06 (.034) &  .06 (.034)  \\
            $3$ & .071 (.045) & .074 (.045) &  .074 (.045) & .074 (.045) & .074 (.045) & .073 (.045)   \\
            $5$ & .1 (.035)  &   .106 (.036) &  .109 (.037)  &  .106 (.036)  &  .106 (.036)  &  .105 (.036)   \\
            $10$ & .135 (.035) &  .16 (.036) &  .169 (.037)  &  .16 (.036) &   .16 (.036)  &  .158  (.036)  \\
        \bottomrule
        \end{tabular}
        \caption{Mean (and standard deviation) RMSE to the simulated conditional probabilities for HiDDeN, logit MLE, logit Bayes, logit GAM, the MLEs for $\bs q$, and the Dirichlet-multinomial (DM) predictive PMF.}
        \label{tab:simulation-results}
\end{table}

\subsection{Discovering Markov Blankets in Synthetic Data} \label{sect:Markov-blanket-sims}

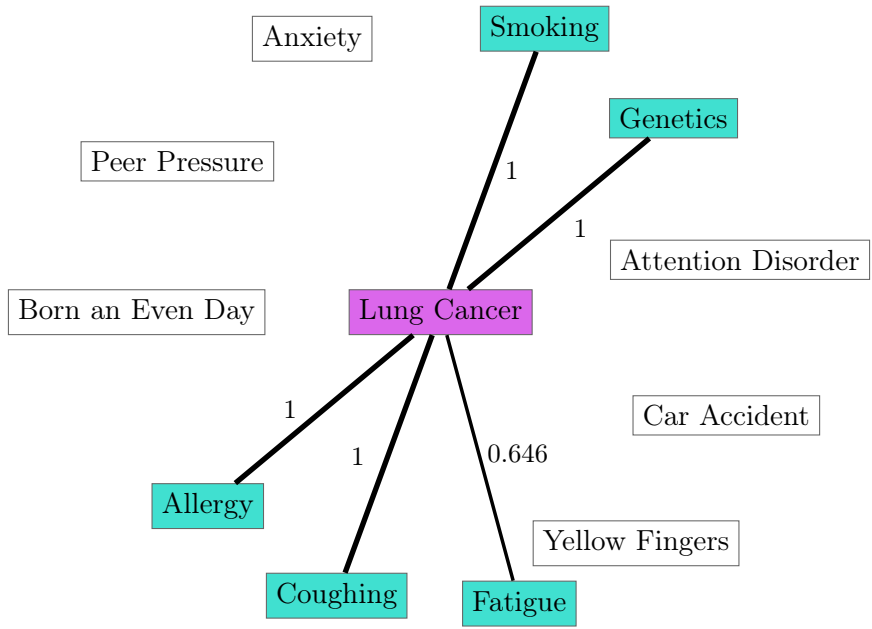
\begin{figure}[t]
    \centering
\begin{tikzpicture}[
    % Node styles
    roundnode/.style={circle, draw=black!60, minimum size=15mm},
    squarednode/.style={rectangle, draw=black!60, minimum size=5mm, fill=#1},
    box/.style = {draw,black,inner sep=10pt,rounded corners=5pt},
    % Edge thickness style with argument
    thick_edge/.style args={#1}{
        line width={0.01pt + #1 * 2.0pt}
    }
]

    % Define custom colors
    \definecolor{turquoise}{HTML}{40E0D0}
    \definecolor{purple}{HTML}{DB67EB}
    \definecolor{whitebg}{HTML}{FFFFFF}

    % Define radius for the circle
    \def\R{4cm} % Radius of the circle for the outer nodes
    
    % Central Node at the origin
    \node[squarednode=purple] (lung_cancer) at (0,0) {Lung Cancer};
    \node[squarednode=whitebg] (anxiety) at (115:\R) {Anxiety};
    \node[squarednode=whitebg] (peer_pressure) at (150:\R) {Peer Pressure};
    \node[squarednode=whitebg] (born_even_day) at (180:\R) {Born an Even Day};
    \node[squarednode=turquoise] (fatigue) at (285:\R) {Fatigue}; % Swapped
    
    \node[squarednode=turquoise] (smoking) at (70:\R) {Smoking};
    \node[squarednode=turquoise] (genetics) at (40:\R) {Genetics};
    
    \node[squarednode=turquoise] (allergy) at (220:\R) {Allergy};
    
    \node[squarednode=whitebg] (attention_disorder) at (10:\R) {Attention Disorder};
    
    \node[squarednode=turquoise] (coughing) at (250:\R) {Coughing};
    \node[squarednode=whitebg] (yellow_fingers) at (310:\R) {Yellow Fingers}; % Swapped
    
    \node[squarednode=whitebg] (car_accident) at (340:\R) {Car Accident};

    % Undirected Edges with weights (Lines)
    \draw[thick_edge=1] (smoking) -- node[pos=0.5, right, font=\small] {1} (lung_cancer);
    \draw[thick_edge=1] (lung_cancer) -- node[pos=0.6, above, font=\small] {1 \: \: \: } (coughing);
    \draw[thick_edge=0.646] (lung_cancer) -- node[pos=0.4, below, font=\small] {\: \: \: \: \: \:  0.646} (fatigue);
    \draw[thick_edge=1] (genetics) -- node[pos=0.6, right, font=\small] {\: \:1} (lung_cancer);
    \draw[thick_edge=1] (allergy) -- node[pos=0.5, left, font=\small] {1 \: } (lung_cancer);
\end{tikzpicture}
    \caption{The HiDDeN MAP estimate for the Markov blanket of lung cancer in the LUCAS data, with edges weighted according to their posterior probability. The true Markov blanket of lung cancer is highlighted in teal.}
    \label{fig:Lung-cancer-Markov-blanket}
\end{figure}

We evaluated HiDDeN's performance in structure learning through application to the LUCAS (Lung Cancer Simple Set) data set for causal discovery \citep{lucas2004bayesian}. An application of HiDDeN to the ALARM network \citep{beinlich1989alarm} is included in the Supplemental material. The LUCAS data are synthetic and intended for modeling the diagnosis and treatment of lung cancer. The network has $p=12$ binary variables (see Figure \ref{fig:Lung-cancer-Markov-blanket}), $|E|=12$ edges, and $n=2000$ observations. We focus on the problem of inferring the Markov blanket for smoking, genetics, fatigue, coughing, allergy, and lung cancer (all of which are presence/absence). Algorithm \ref{alg:edge-indicators} is run to select a parent set from $2^{p-1}$ possibilities for each of the six variables, with the number of iterations equal to $10,000$, $200$ of which are discarded as burn-in, and the stepsizes chosen according to the acceptance probabilities of the MALA-within-Gibbs sampler and MCMC diagnostics. HiDDeN is compared to group lasso, and the grow-shrink (GS) and Incremental Association (IAMB) structure learning algorithms \citep{scutari2010learning}. The Hamming distances from the true Markov blanket are shown in Table \ref{tab:Lucas}. 

\begin{figure}
    \centering
    \includegraphics[width=0.9\linewidth]{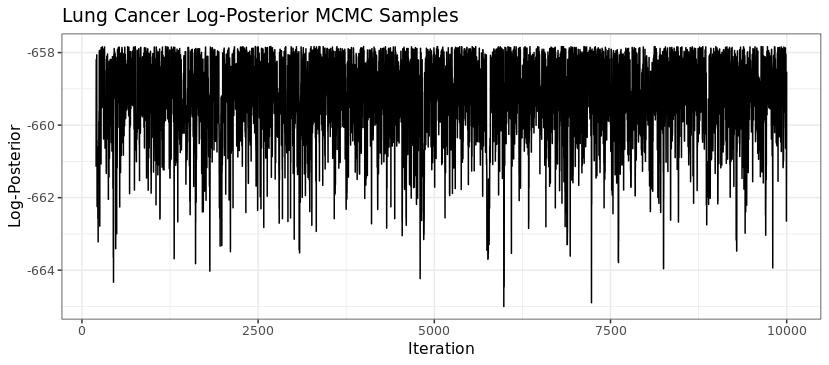}
    \caption{Traceplot for the log-posterior $\log f(\bs t_j, \bs z_j \mid \bs n)$ for the lung cancer variable in the LUCAS dataset.}
    \label{fig:lucas-traces}
\end{figure}

The HiDDeN MAP estimate is equivalent to the true Markov blanket for all variables, with the exception of smoking; this leads to HiDDeN achieving the lowest total Hamming distance. The original aim of the LUCAS dataset is to discover the Markov blanket for lung cancer \citep{lucas2004bayesian}; therefore, we provide an additional discussion on this specific variable. Figure \ref{fig:Lung-cancer-Markov-blanket} shows the HiDDeN MAP estimate with edges weighted according to their posterior probability. We are uncertain about the association with fatigue, as this variable has a posterior probability of $0.646.$ Regarding HiDDeN's computation, the acceptance probabilities of $\bs t_j(1)$ and $\bs t_j(2)$ after burn-in are $0.557$ and $0.560$, leading to an effective sample size (ESS) of $1423.543$ for the log-posterior. In addition, the traceplot for the log-posterior in Figure \ref{fig:lucas-traces} suggests convergence of the MCMC chain.

\begin{table}[t]
\centering
\begin{tabular}{rrrrr}
  \toprule
 Variable & HiDDeN & Group lasso & GS & IAMB \\ 
  \midrule
Smoking & 2 & 1 & 1 & 2 \\ 
  Genetics & 0 & 2 & 0 & 1 \\ 
  Fatigue & 0 & 0 & 2 & 0 \\ 
  Coughing & 0 & 0 & 0 & 0 \\ 
  Allergy & 0 & 0 & 0 & 1 \\ 
  Lung Cancer & 0 & 4 & 0 & 0 \\ 
   \midrule
   Total & 2 & 7 & 3 &  4 \\
   \bottomrule
\end{tabular}
\caption{Hamming distances to the true Markov blankets for six variables in the LUCAS dataset.}
\label{tab:Lucas}
\end{table}

\subsection{Evaluating Two DAGs}

We now consider the problem of selecting two DAGs. Let $G_m = ([3],E_m)$ for $m=1,2$, where $E_1 = \{ (1,2), (1,3) \}$ and $E_2 = \{ (1,2), (1,3), (2,3) \}$ represent two possible DBNs for $p=3$ variables.  The variables can be interpreted as a set of confounders, $x_1$, a treatment or intervention, $x_2$, and an outcome, $x_3$. Selecting one of the two DAGs is equivalent to assessing the impact of $x_2$ on $x_3$ conditional on $x_1$, i.e. whether an intervention has an effect on the outcome after adjusting for all confounders. We assume that $k_2=k_3=2$ and simulate data according to
\begin{align*}
    \Pr(x_1)& =1/k_1; & \Pr(x_2 \mid x_1) & \sim \tx{Unif}(0.01, 0.99); & \Pr(x_3 \mid x_1,x_2) & \sim  \tx{Beta}(2,15); & \forall & x_1,x_2,x_3.
\end{align*}
That is, we are generating data from $G_2$. We fix the sample size at $n=200$, then vary $k_1 \in \{ 5, 25, 100, 200 \}$ to mimic increasing the dimension and/or the number of categories in the confounder variables. In turn, this increases the dimension of the contingency tables $\bs n_{\Pa(j),j}$. We apply HiDDeN with Algorithm \ref{alg:unknown-DAG} and $G_1$ and $G_2$ as candidate graphs, with prior probabilities $\Pr(G_m)=1/2$. The MALA-within-Gibbs algorithm is run for $10000$ iterations with $200$ burn-in samples. For $100$ independent replications, we record the MAP estimate $\widehat G$, as well as the graph selected from $\{ G_1,G_2 \}$ using the Bayesian information criterion (BIC), Akaike information criterion (AIC), and BDE score. 

The proportion of correctly selected DAGs are detailed in Table \ref{tab:correct-DAG-choices}. HiDDeN is the only method that selects $G_2$ in more than half of the replications for any value of $k_1$. AIC and BIC completely fail at selecting $G_2$ for $k_2>5$, preferring the model with fewer parameters. Interestingly, both of the methods derived from the Dirichlet prior benefit from increasing $k_1$, a consequence of the reliance of the marginal likelihood function in \eqref{eq:n-given-pi-tilde} on $k_1$. However, at its most accurate, BDE still concludes in favor of the incorrect DAG in more than half of the replications, reducing the choice to a coin flip. An additional example comparing HiDDeN with AIC, BIC, and BDE is included in the Supplemental Material.

\begin{table}[t]
\centering
\begin{tabular}{rrrrr}
  \toprule
  $k_1$ & 5 & 25 & 100 & 200 \\ 
  \midrule
HiDDeN & 0.61 & 0.71 & 0.76 & 0.82 \\ 
  BIC & 0.02 & 0.00 & 0.00 & 0.00 \\
  AIC & 0.38 & 0.00 & 0.00 & 0.00 \\ 
  BDE & 0.08 & 0.01 & 0.21 & 0.49 \\
   \bottomrule
\end{tabular}
\caption{Proportion of simulations in which $G_2$ is selected for HiDDeN, BIC, AIC, and BDE.}
\label{tab:correct-DAG-choices}
\end{table}

\section{The METABRIC Dataset} \label{sect:METABRIC}

\begin{table}
    \centering
    \begin{tabular}{cccc}
    \toprule
        Name & Description & Categories & $k_j$  \\
    \midrule
        TBS & Type of breast surgery & Mastectomy, breast conserving & 2   \\
         CHT & Chemotherapy &    Yes,no (received chemotherapy) & 2   \\
         DFC & Death from cancer & Died of disease, died of other causes, living & 3   \\
    \midrule
         CTD & Cancer type detailed & IDC, mDLC, ILC, MMBC, MpBC &  5  \\
         AAD & Age at diagnosis &  [20,30), [30,40) \dots, [90,$\infty$) & 8   \\
         TSI & Tumor size  &  Low, normal, high & 3   \\
         RTH & Radio therapy  &  Yes,no (received radio therapy) & 2  \\
         PTL & Primary tumor laterality  & Right breast, left breast & 2   \\
         INC & Molecular subtype of cancer &  $1,2,\dots, 4\tx{ER}+,4\tx{ER}-,5, \dots, 9$ & 10 \\
         IMS & Inferred menopausal state  & Post,pre & 2   \\
         HTH & Hormone therapy  &  Yes,no (received hormone therapy) & 2   \\
        COH & Cohort  &  Group $1$, $2$, $3$, $4$, $5$ & 5 \\ 
    \bottomrule
    \end{tabular}
    \caption{Name, description, categories, and number of categories for the METABRIC variables. The three outcomes are listed first, then the covariates.}
    \label{tab:metabric-variables}
\end{table}

Despite improved outcomes due to screening and adjuvant therapies \citep{hashim2016global}, breast cancer remains an urgent problem: \cite{sung2021global} estimated that there were 2.3 million new cases and more than 685,000 deaths in 2020. Moreover, the incidence of breast cancer depends on a complicated array of genetic and non-genetic factors \citep{nolan2023deciphering}.  Treatment of breast cancer consists of preoperative systemic therapy, surgery, and postoperative therapy, depending on stage \citep{trayes2021breast}. We apply HiDDeN to the Molecular Taxonomy of Breast Cancer International Consortium (METABRIC) dataset \citep{pereira2016somatic} to explore the relationships between factors and outcomes in breast cancer treatment. The METABRIC data consist of 31 clinical attributes, z-scores for m-RNA levels in 331 genes, and mutation for 175 genes from $1,980$ breast cancer samples. We focus on the attributes in Table \ref{tab:metabric-variables}; most of the variables are clinical with the exception of integrative cluster (INC), defined as the molecular cancer subtype of a sample derived from gene expression. Our goal is to infer the network structure of the following outcomes given covariates: breast surgery type (TBS), chemotherapy (CHT), and death from cancer (DFC). TBS and CHT are binary, whereas DFC has three levels (died from cancer, died from other causes, living). All covariates in Table \ref{tab:metabric-variables} are included in the candidate set for each outcome, these consist of characteristics of the tumor and patient, as well as additional treatments.

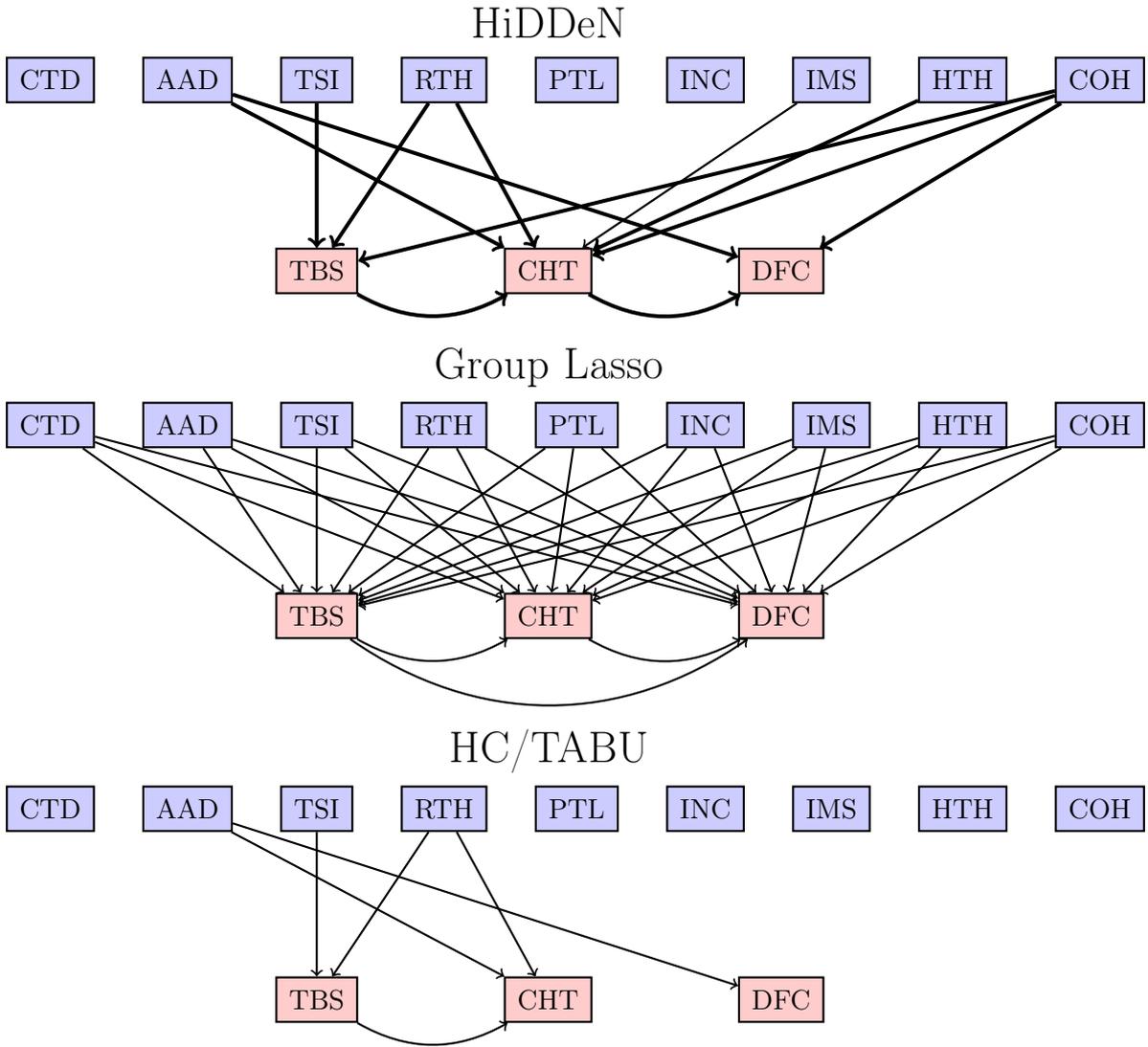
\begin{figure}[t]
\centering

\begin{tikzpicture}[
node distance=.5cm,
every node/.style={
draw,
thick,
align=center,
inner sep=5pt
},
covariate/.style={fill=blue!20},
outcome/.style={fill=red!20},
every path/.style={
->,
thick
},
thick_edge/.style args={#1}{
        line width={0.01pt + #1 * 1.5pt}
}
]
\node[font=\LARGE, draw=none] at (6.85, .8) {HiDDeN};
% Define Covariate nodes on a single horizontal line
\node[covariate] (ctd) at (0, 0) {CTD};
\node[covariate] (aad) [right=0.65cm of ctd] {AAD};
\node[covariate] (tsi) [right=0.65cm of aad] {TSI};
\node[covariate] (rth) [right=0.65cm of tsi] {RTH};
\node[covariate] (ptl) [right=0.65cm of rth] {PTL};
\node[covariate] (inc) [right=0.65cm of ptl] {INC};
\node[covariate] (ims) [right=0.65cm of inc] {IMS};
\node[covariate] (hth) [right=0.65cm of ims] {HTH};
\node[covariate] (coh) [right=0.65cm of hth] {COH};
% Define Outcome nodes on a single horizontal line, in the new order
\node[outcome] (tbs) [below=2cm of tsi] {TBS};
\node[outcome] (cht) [right=2cm of tbs] {CHT};
\node[outcome] (dfc) [right=2cm of cht] {DFC};
% Draw edges
\draw[thick_edge=1] (tsi) -> (tbs);
\draw[thick_edge=1] (rth) -> (tbs);
\draw[thick_edge=.989] (coh) -> (tbs);
\draw[thick_edge=1] (aad) -> (cht);
\draw[thick_edge=1] (rth) -> (cht);
\draw[thick_edge=.543] (ims) -> (cht);
\draw[thick_edge=1] (hth) -> (cht);
\draw[thick_edge=.999] (coh) -> (cht);
% Draw the curved outcome-to-outcome edges
\draw[thick_edge=.999] (tbs) to [bend right=30] (cht);
\draw[thick_edge=1] (cht) to [bend right=30] (dfc);
% Draw other edges to DFC
\draw[thick_edge=1] (aad) -> (dfc);
\draw[thick_edge=1] (coh) -> (dfc);
\end{tikzpicture}

\vspace{.1cm}

\begin{tikzpicture}[
node distance=.5cm,
every node/.style={
draw,
thick,
align=center,
inner sep=5pt
},
covariate/.style={fill=blue!20},
outcome/.style={fill=red!20},
every path/.style={
->,
thick
}
]
\node[font=\LARGE, draw=none] at (6.85, .8) {HC/TABU};
% Define Covariate nodes on a single horizontal line
\node[covariate] (ctd) at (0, 0) {CTD};
\node[covariate] (aad) [right=0.65cm of ctd] {AAD};
\node[covariate] (tsi) [right=0.65cm of aad] {TSI};
\node[covariate] (rth) [right=0.65cm of tsi] {RTH};
\node[covariate] (ptl) [right=0.65cm of rth] {PTL};
\node[covariate] (inc) [right=0.65cm of ptl] {INC};
\node[covariate] (ims) [right=0.65cm of inc] {IMS};
\node[covariate] (hth) [right=0.65cm of ims] {HTH};
\node[covariate] (coh) [right=0.65cm of hth] {COH};
% Define Outcome nodes on a single horizontal line, in the new order
\node[outcome] (tbs) [below=2cm of tsi] {TBS};
\node[outcome] (cht) [right=2cm of tbs] {CHT};
\node[outcome] (dfc) [right=2cm of cht] {DFC};
% Draw a directed edge from every covariate to every outcome
\draw (tsi) to (tbs);
\draw (rth) to (tbs);
\draw (aad) to (cht);
\draw (rth) to (cht);
\draw (aad) to (dfc);
\draw(tbs) to [bend right=30] (cht);

\end{tikzpicture}

\vspace{-0.1cm}
\begin{tikzpicture}[
node distance=.5cm,
every node/.style={
draw,
thick,
align=center,
inner sep=5pt
},
covariate/.style={fill=blue!20},
outcome/.style={fill=red!20},
every path/.style={
->,
thick
}
]
\node[font=\LARGE, draw=none] at (6.85, .8) {Penalized MLR};
% Define Covariate nodes on a single horizontal line
\node[covariate] (ctd) at (0, 0) {CTD};
\node[covariate] (aad) [right=0.65cm of ctd] {AAD};
\node[covariate] (tsi) [right=0.65cm of aad] {TSI};
\node[covariate] (rth) [right=0.65cm of tsi] {RTH};
\node[covariate] (ptl) [right=0.65cm of rth] {PTL};
\node[covariate] (inc) [right=0.65cm of ptl] {INC};
\node[covariate] (ims) [right=0.65cm of inc] {IMS};
\node[covariate] (hth) [right=0.65cm of ims] {HTH};
\node[covariate] (coh) [right=0.65cm of hth] {COH};
% Define Outcome nodes on a single horizontal line, in the new order
\node[outcome] (tbs) [below=2cm of tsi] {TBS};
\node[outcome] (cht) [right=2cm of tbs] {CHT};
\node[outcome] (dfc) [right=2cm of cht] {DFC};
% Draw a directed edge from every covariate to every outcome
\foreach \covariate in {ctd, aad, tsi, rth, ptl, inc, ims, hth, coh} {
\draw (\covariate) -> (tbs);
\draw (\covariate) -> (cht);
\draw (\covariate) -> (dfc);
}
% Draw the curved outcome-to-outcome edges
\draw (tbs) to [bend right=30] (cht);
\draw (cht) to [bend right=30] (dfc);
\draw (tbs) to [bend right=35] (dfc);
\end{tikzpicture}

\caption{Median probability model for the METABRIC variables after fitting HiDDeN (top), HC and TABU algorithms (bottom), and penalized MLR (bottom). Covariates are displayed in blue and outcomes in red. Line thickness in the top DAG is weighted by posterior edge probability.}
\label{fig:METABRIC-DAG}
\end{figure}

We assume the outcomes have a stochastic ordering of TBS, CHT, and DFC; implying that CHT is in the candidate set of TBS, and both CHT and TBS are in the candidate set of DFC. This ordering is based on the typical ordering of these events, though chemotherapy is sometimes performed pre-surgery. The data have $n=1,867$ samples with $p=9$ variables ($3$ outcomes and $8$ covariates). We apply Algorithm \ref{alg:edge-indicators} and fit HiDDeN for $10,000$ iterations with the initial $500$ removed for burn-in. The median probability DAG is employed as a point estimate for HiDDeN. For comparison, we implement the hill climbing (HC) and TABU structure learning algorithms to each outcome. We also train a multinomial logistic regression with group lasso penalty (penalized MLR) sequentially for TBS, CHT, DFC. We use group lasso to penalize coefficients across the categories of each parent variable. The selected penalty is the largest within 1 standard error of the minimum mean cross-validated error. We define the parents of any node to be those with a non-zero coefficient.

\begin{figure}[t]
    \centering
    \includegraphics[scale=0.65]{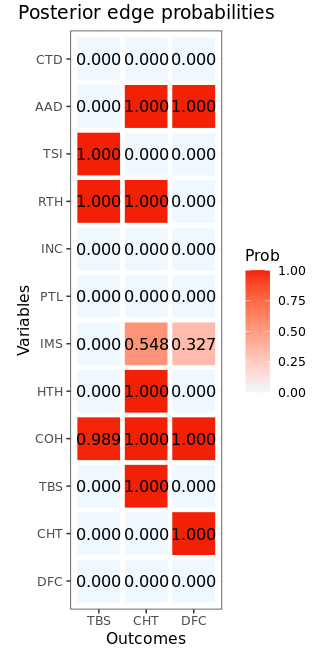}
    \caption{Posterior edge probabilities for the METABRIC dataset. The $y$ axis displays parents, while the $x$ axis displays the outcome variables of interest.}
    \label{fig:METABRIC-posterior-probs}
\end{figure}

The DAG estimates produced by all methods are shown in Figure \ref{fig:METABRIC-DAG}; the posterior edge probabilities for HiDDeN are plotted in Figure \ref{fig:METABRIC-posterior-probs}. The median probability DAG provides interesting insights into the conditional dependence structure of the outcomes. For example, the parent set of TBS includes tumor size (TSI), radiotherapy (RTH), and cohort (COH). The largest parent set is that of CHT, which includes age at diagnosis (AAD), RTH, inferred menopausal state (IMS), hormone therapy (HTH), and COH. Finally, DFC has parents AAD, COH, and CHT. HC and TABU infer an the same DAG, which has a sparser structure than that inferred by HiDDeN. The penalized MLR selects the entire candidate set as the parent set, resulting in a connected DAG. The same behavior holds when the penalty is chosen to minimize the mean cross-validated error. 

Figure \ref{fig:METABRIC-DAG} shows that conditional on all other variables, mortality only depends on the age of the patient at diagnosis, whether or not they received chemotherapy, and their cohort. Hence, these variables make the DFC independent of tumor characteristics (TSI), the administration of other therapies (TBS, RTH, and HTH), the allocation of the sample into a molecular subtype (INC) and the type of cancer (CTD). In fact, of the three variables that separate the samples into groups based off shared characteristics (CTD, INC, and COH), only COH is included in the parent sets of the outcomes. Figure \ref{fig:METABRIC-posterior-probs} conveys uncertainty in the edges between variables. Several of the edges are allocated posterior probability close to either $0$ or $1$. However, the menopausal state of the patient has an uncertain status in the parent sets: the edges (IMS, CHT) and (IMS, DFC) have posterior mass $0.548$ and $0.327$, respectively. Only the former edge is included in the median probability DAG in Figure \ref{fig:METABRIC-DAG}. 

\section{Discussion} \label{section:discussion}

This article proposes a new Bayesian method for parameter and structure learning in DBNs. We specify a hierarchical Dirichlet model for the conditional probabilities of a node, $j$, given the categories of its parents, $\Pa(j)$. A key to our approach is learning latent concentrations, $\bs t_j$, adaptively from the data. We derive a novel marginal MALA-within-Gibbs sampler to generate posterior samples of $\bs t_j$. Our computation strategy is motivated by a key property: the full conditional distribution of the latent concentrations is factorizable and almost-surely log-concave. We generalize our approach to infer parent sets, Markov blankets, and DAGs, then construct appropriate estimators and uncertainty quantification metrics. In several simulation studies, we demonstrate the strong performance of our method in parameter estimation, structure learning, and graph selection. Finally, our approach infers insightful networks describing the relationships between genetic and non-genetic factors with various outcomes in breast cancer. 

There are many possible extensions to our work. For instance, existing adaptive algorithms for selecting $\bs \epsilon_j$ \citep{atchade2006adaptive, marshall2012adaptive, biron2024auto} could be applied to our MCMC scheme and circumvent tuning. More generally, gradient based MCMC methods, such as Hamiltonian Monte Carlo (HMC) or the No-U-Turn Sampler (NUTS) \citep{hoffman2014nouturn}, could be used to sample from the posterior of $\bs t_j$. In addition, other priors besides the gamma distribution may be considered for $\bs t_j(x_j)$, though some choices do not retain the log-concavity property. With the gamma prior, our MALA-within-Gibbs algorithm can be used for posterior inference in finite approximations to the HDP, and could be adapted to address extensions of the HDP \citep{catalano2025hierarchical}, though log-concavity will fail if there are empty mixture components. We propose several algorithms to obtain MCMC samples of graphical structures, but HiDDeN can be adjusted to incorporate other MCMC DAG algorithms; see \cite{liang2023structure} and references therein. 

A promising area of future research is to extend HiDDeN to different settings than those considered in this article, such as weighted Bayesian networks and undirected graphs. With regard to the latter models, our approach can infer \textit{local} undirected graphs via the Markov blankets, thus our strategy could be extended to model an entire undirected graph. Alternatively, we can construct a single undirected graph by iteratively moving from node-to-node, fitting HiDDeN, and keeping nodes in the candidate set based on the point estimate, say, the median probability model, from the previous node. In a broader sense, HiDDeN can be used to infer network structure in latent discrete variables, such as those in Bayesian pyramids \citep{gu2023bayesian} and discrete hierarchical models \citep{kong2024learning}. This motivates using HiDDeN in a deep generative model for categorical data. As an example, if we start with a variable $j$ and its parent variables $\bs x_{\Pa(j)}$, we could model the conditional cell counts in the following multi-level structure. Let $\bs y_j = (y_{j1} \dots, y_{jL})$ be latent variables with $y_{jl} \in [k_y]$, where the conditional probabilities are of each $y_{jl}$ are modeled according to the HiDDeN prior. Sampling the latent variables creates latent counts, say, $\bs n_{\bs y_j}$. From these latent counts, we then generate $x_j$ via HiDDeN, i.e. we set $\bs n_{\Pa(j)} = \bs n_{\bs y_j}$ in our notation. This process effectively creates a latent layer between $\Pa(j)$ and $j$, in which unobserved nodes $1, \dots, L$ characterize the dependence structure. It is straightforward to see how this setup could be extended to multiple layers of latent variables, resulting in a novel multi-layer model for discrete data.

\section*{Acknowledgments}

This work was supported by the National Institutes of Health (NIH) under Grants 1R01AI155733, 1R01ES035625, and 5R01ES027498; the United States Office of Naval Research (ONR) Grant N00014-24-1-2626; the European Research Council (ERC) under Grant 856506; and Merck \& Co., Inc., through its support for the Merck BARDS Academic Collaboration. Dombowsky was also funded by the Myra and William Waldo Boone Fellowship.

\section*{Supplementary Material}

The Supplementary Material contains proofs, extended technical details on the MALA-within-Gibbs sampler, additional comments on the simulation studies and application, two extra simulation studies, a method for computing maximum marginal likelihood estimators in the HiDDeN model, and a derivation of the Bayes factor for HiDDeN under certain conditions on the prior. Code to reproduce all results can be found at \url{github.com/adombowsky/HiDDeN_DAG}.

\bibliography{sources}

\appendix

\section*{Supplementary Material}

\section{Proofs}

\subsection{Proof of Proposition \ref{prop:poisson-multinomial}}
\begin{proof}

Observe that we construct the likelihood by sampling $\bs n_{j \mid \Pa(j)} \mid \bs n_{\Pa(j)}$ via independent multinomial samples, where
\begin{equation} \label{eq:SM-conditional-multinomial}
\begin{gathered}
    [\bs n_{\Pa(j),j}(\bs x_{\Pa(j)},1), \dots, \bs n_{\Pa(j),j}(\bs x_{\Pa(j)},k_j)] \mid \bs n_{\Pa(j)}(\bs x_{\Pa(j)})  \\
    \sim \tx{Mult}(\bs n_{\Pa(j)}(\bs x_{\Pa(j)}), [\bs \pi_{j \mid \Pa(j)}(1 \mid \bs x_{\Pa(j)}, \dots,  \pi_{j \mid \Pa(j)}(
    k_j\mid \bs x_{\Pa(j)}])
\end{gathered}
\end{equation}
We have that
\begin{equation*}
    \bs n_j(x_j) = \sum_{\bs x_{\Pa(j)}} \bs n_{\Pa(j),j}(\bs x_{\Pa(j)},x_j) = \sum_{\bs x_{\Pa(j)}}\sum_{i:\bs x_{i \Pa(j)} = \bs x_{\Pa(j)}}z_{ij},
\end{equation*}
where $z_{ij} = \textbf{1}_{x_{ij} = x_j}$. Under \eqref{eq:SM-conditional-multinomial}, each $z_{ij}$ is independent, with $\Pr(z_{ij}=1 \mid \bs \pi) = \bs \pi_{j \mid \Pa(j)}(x_j \mid \bs x_{\Pa(j)})$. Therefore, the vector $\bs n_j = (\bs n_j(1), \dots, \bs n_j(x_j))$ is distributed as a Poisson-multinomial with success probabilities $\bs \pi_{j \mid \Pa(j)}(\cdot \mid \bs x_{i\Pa(j)}).$
    
\end{proof}

\subsection{Proof of Theorem \ref{thm:log-concave}}

\begin{proof}
    Observe that
\begin{gather*}
    \log h_{j,x_j}(t) \propto -b_j t + (\rho_j/k_j -1) \log t 
    \\ + \sum_{x_{\Pa(j)}} \{  \log \Gamma(\bs n_{\Pa(j),j}(\bs x_{\Pa(j)}) + t) - \log \Gamma(t) + t \log \bs u_j(\bs x_{\Pa(j)}) \}.
\end{gather*}
Hence,
\begin{gather}
    \frac{d \log h_{j,x_j}(t)}{dt} = -b_j + \frac{\rho_j/k_j-1}{t} + \sum_{\bs x_{\Pa(j)}} \{ \psi(\bs n_{\Pa(j),j}(\bs x_{\Pa(j)}, x_j) + t) - \psi(t) + \log \bs u_j(\bs x_{\Pa(j)}) \} \label{eq:SM-t-gradient} \\
    \implies \frac{d^2 \log h_{j,x_j}(t)}{dt^2} = \frac{1 - \rho_j/k_j}{t^2} + \sum_{\bs x_{\Pa(j)}} \{ \psi^{(1)}(\bs n_{\Pa(j),j}(\bs x_{\Pa(j)}, x_j) + t) - \psi^{(t)}(t) \}. \label{eq:SM-t-hessian}
\end{gather}
If $\rho_j \geq k_j$ and $\bs n_{j}(x_j)>0$, then $\frac{d^2 \log h_{j,x_j}(t)}{dt^2} <0$, as the trigamma function $\psi^{(1)}(t)$ is decreasing for all $t>0$.
\end{proof}

\section{MALA-Within-Gibbs Sampler}

In this section, we discuss the technical details for the MALA-within-Gibbs sampler that we propose for HiDDeN. The initial value $\bs t_j^{(0)}(x_j)$ must be supplied for the algorithm. Too small a value of $\bs t_j^{(0)}(x_j)$ often leads to low acceptance probabilities. In practice, we find that setting $\bs t_j^{(0)}(x_j) \geq 1$ avoids this pathology. For subsequent iterations, we generate a candidate $\bs t_j^*(x_j)$ from the current value $\bs t_j^{(r)}(x_j)$, then accept or reject that candidate. The Langevin proposal is
\begin{equation} \label{eq:SM-langevin-proposal}
    \bs t_j^*(x_j) = \bs t_j^{(r)}(x_j) + \frac{\bs \epsilon_j(x_j)^2}{2}  \frac{d \log h_{j,x_j}(\bs t_j^{(r)}(x_j))}{dt} + \bs \epsilon_j(x_j) z^{(r)}
\end{equation}
where $z^{(r)} \sim \N(0,1)$ and the derivative of $h_{j,x_j}$ is defined in \eqref{eq:SM-t-gradient}. The gradient will depend on the current value of $\bs u_j$. We will then set $\bs t_j^{(r+1)}(x_j) = \bs t_j^*(x_j)$ with probability
\begin{equation*}
   \bs \kappa_j^{(r+1)}(x_j) = \min \lb 1, \frac{h_{j,x_j}(\bs t_j^{*}(x_j)) \psi_{j,x_j}(\bs t_j(x_j)^{(r)}; \bs t_j(x_j)^*) }{h_{j,x_j}(\bs t_j^{(r)}(x_j))\psi_{j,x_j}(\bs t_j(x_j)^*; \bs t_j(x_j)^{(r)})} \rb,
\end{equation*}
where
\begin{equation*}
    \psi_{j,x_j}(t^*;t) = \phi\left(t^*; t + \frac{\bs \epsilon_j(x_j)^2}{2} \frac{d \log h_{j,x_j}(t)}{dt}, \bs \epsilon_j(x_j)^2 \right).
\end{equation*}
Otherwise, we set $\bs t_j^{(r+1)}(x_j) = \bs t_j^{(r)}(x_j) $. Any proposals from \eqref{eq:SM-langevin-proposal} that violate the positivity constraint on $\bs t_j(x_j)$ will automatically set $\bs \kappa_j^{(r+1)}=0$ and $\bs t_j^{(r+1)}(x_j) = \bs t_j^{(r)}(x_j)$. Hence, the user should be careful not to set $\bs \epsilon_j(x_j)$ to be too large, as this choice will likely cause more rejections. We also record the acceptance indicator of the iterate, $\textbf{1}_{\bs t_j^{(r+1)}(x_j) = \bs t_j^{(r)}(x_j)}$ for all $r>0$. The acceptance proportion of the MCMC samples is $(1/R)\sum_{r}\textbf{1}_{\bs t_j^{(r+1)}(x_j) = \bs t_j^{(r)}(x_j)}$, which is used to select $\bs \epsilon_j(x_j)$, e.g., by aiming for the asymptotically optimal rate of $0.574$ \citep{roberts1998optimal}. MCMC pathologies are monitored by computing samples from the log-posterior, which has the form
\begin{gather*}
    \log f(\bs t_j, \bs z_j \mid \bs n) \propto \sum_{\bs x_{\Pa(j)} \in \chi_{\Pa(j)}} \lb \log \Gamma(\beta_j) - \log \Gamma(\bs n_{\Pa(j)}(\bs x_{\Pa(j)}) + \beta_j) \rb \\
    + \sum_{x_j=1}^{k_j} \lb - b_j \bs t_j(x_j) +  (\rho_j/k_j - 1) \log \bs t_j(x_j) - K_{\Pa(j)} \log \Gamma(\bs t_j(x_j))  \rb \\
    + \sum_{x_j,\bs x_{\Pa(j)}} \log \Gamma(\bs n_{\Pa(j),j}(\bs x_{\Pa(j)}, x_j) + \bs t_j(x_j)) \\ 
    + \log \Gamma \bigg ( \sum_{j^\prime \in \Ca(j)} z_{j^\prime j} + c_j\bigg ) + \log \Gamma \bigg ( \sum_{j^\prime \in \Ca(j)} (1-z_{j^\prime j}) + d_j \bigg ). 
\end{gather*}

\section{Experiments}

\subsection{Parameter Learning for Sparse Counts}
Recall that we are interested in estimating $\bs q = (q_{x_{\Pa(j)}})_{x_{\Pa(j)} \in [K_j]}$, the vector of conditional probabilities for all parent categories. For all values of $K_{\Pa(j)}$, we fix $\bs \epsilon_j = c(0.5,0.5)$, $b_j=1$, and $\rho_j=2$. We use \texttt{rstanarm} \citep{rstanarm}, \texttt{mgcv} \citep{wood2011fast}, and \texttt{bnlearn} \citep{scutari2010learning} to implement the Bayesian logistic regression, GAM logistic regression, and Dirichlet-multinomial, respectively. All hyperparameters for competitors are set to their defaults.

\subsection{Discovering Markov Blankets in Synthetic Data}

\begin{table}[t]
    \centering
    \begin{tabular}{cc}
    \toprule
        Variable & $\tx{Mb}(j)$ \\
    \midrule
        smoking & anxiety, peer pressure, yellow fingers, lunc cancer, genetics \\
        genetics & lung cancer, attention disorder, smoking \\
        fatigue & lung cancer, coughing, car accident, attention disorder \\
        coughing & allergy, lung cancer, fatigue \\
        allergy & coughing, lung cancer \\
        lung cancer & smoking, genetics, fatigue, coughing, allergy \\
    \bottomrule
    \end{tabular}
    \caption{Parent sets for the LUCAS data.}
    \label{tab:lucas-parent-sets}
\end{table}

\begin{table}
    \centering
    \begin{tabular}{cccc}
    \toprule
        Variable & Stepsize  & Acceptance Probabilities & $\log f(\bs t_j, z_j \mid \bs n)$ ESS \\
    \midrule
        smoking & (.30, .30) & (.554, .534) & 1311.484 \\
        genetics & (.47, .18)  & (.779, .691)  & 329.891 \\
        fatigue & (.32, .32) & (.564, .650)  & 1828.669 \\
        coughing & (.20, .33) & (.611, .602)  & 1086.679  \\
        allergy & (.65, .35) & (.524, .516) & 2149.449  \\
        lung cancer & (.11, .15) & (.557, .560)   &  1423.543\\
    \bottomrule
    \end{tabular}
    \caption{Stepsize, acceptance probability, and effective sample size (ESS) of the log-posterior when applying HiDDeN to the LUCAS data.}
    \label{tab:lucas-info}
\end{table}

In the context of DBNs, the Markov blanket, denoted $\tx{Mb}(j)$, of a node $j$ is the union of $\Pa(j)$, its children, and the parent sets of its children \citep{pearl1988probabilistic}. The LUCAS data are \href{https://www.causality.inf.ethz.ch/data/LUCAS.html}{publicly available}, and the true Markov blankets for each variable are listed in Table \ref{tab:lucas-parent-sets}. We encode $\tx{Mb}(j)$ from Table \ref{tab:lucas-parent-sets} as a binary vector $\bs z_j^0 \in \{ 0,1 \}^{p-1}$. In the main article, we report the Hamming distance $\tx{H}(\widehat{\bs z}_j,\bs z_j^0)$ from an estimator $\widehat{\bs z}_j$ to the true Markov blanket.

We do not preprocess the data. The hyperparameters for HiDDeN are $b_j=1$, $\rho_j=k_j+1$, $c_j=d_j=1$ for all variables, and we initialize the concentrations at $\bs t_j = (1,1)$. MALA stepsizes are tuned on preliminary runs, initializing at $\bs \epsilon_j = (.1,.1)$, then steadily increasing these values. We inspect the acceptance probabilities and traceplots for $\bs t_j$ and the log-posterior, to ensure that the chain is (a) not getting stuck in a region due to a high number of rejections, and (b) sufficiently exploring the parameter space. The stepsizes, acceptance probabilities, and effective sample sizes (ESS) for $\log f(\bs t_j, \bs z_j \mid \bs n)$ across the variables are detailed in Table \ref{tab:lucas-info}. Group lasso is implemented in \texttt{glmnet} using the multinomial likelihood (setting \texttt{family="multinomial"} and \texttt{type.multinomial="grouped"}). The penalty is selected using cross validation, then choosing the largest value within 1 standard error of the minimum mean cross-validated error (\texttt{lambda.1se}). The \texttt{bnlearn} methods are applied using the default settings and hyperparameters in \texttt{R}. 

\subsection{Evaluating Two DAGs} 
The setup of this simulation study can arise when one wants to evaluate whether there is an association between $x_2$ and $x_3$ conditional on $x_1$, i.e. testing $H_0: x_2 \ind x_3 \mid x_1$. We assume that the dependence between $x_1$ and $(x_2,x_3)$ is known to the statistician, such as when $x_1$ is a set of confounders for two variables. Hence, we propose two possible DAGs, $G_m$ for $m=1,2$. Both DAGs include the edges $(1,2)$ and $(1,3)$, but $G_2$ also includes the edge $(2,3)$, i.e., an association between $x_2$ and $x_3$ conditional on $x_1$. For HiDDeN, we evaluate the DAGs by choosing the $G_m$ that maximizes the posterior probability. Recall that the posterior probability of $G_m$ is
\begin{equation*}
    \Pr(G_m \mid \bs n) = \int_{\bs t_3} \frac{\Pr(G_m) f(\bs n_{3 \mid \Pa(3)_m} \mid \bs n_{\Pa(3)_m}, \bs t_3)}{\sum_{m^\prime = 1}^{2} \Pr(G_{m^\prime})f(\bs n_{3 \mid \Pa(3)_{m^\prime}} \mid \bs n_{\Pa(3)_{m^\prime}}, \bs t_3) } \tx{d} f(\bs t_3 \mid \bs n).
\end{equation*}
The term in the numerator is the exponentiated BDE of $G_m$ with $\bs t_j$ as the prior sample size, motivating a comparison to the standard BDE, as well the AIC and BIC, all of which are commonly used as DAG scores \citep{kitson2023survey}.

For the values of the confounder categories $k_1$, we use stepsizes $(.9,.3)$, $(.65,.25)$, $(.45,.15)$, $(.4,.1)$, respectively, hyperparameters $b_3=1$ and $\rho_3=2$, and initializations $\bs t_3^{(0)} = (1,1)$. These values are constant across the $100$ replications. BIC, AIC, and BDE are implemented with the \texttt{bnlearn} package and evaluated at both $G_1$ and $G_2$, and all score hyperparameters are set to their defaults. For the competitors, we set the estimator of $G$ to be the $G_m$ with the highest score, e.g., for BDE this rule is: $\widehat G = G_2$ if and only if $\tx{BDE}(G_2)-\tx{BDE}(G_1)>0$, otherwise $\widehat G = G_1$.

\section{METABRIC Dataset}

\begin{figure}
    \centering
    \includegraphics[scale=0.61]{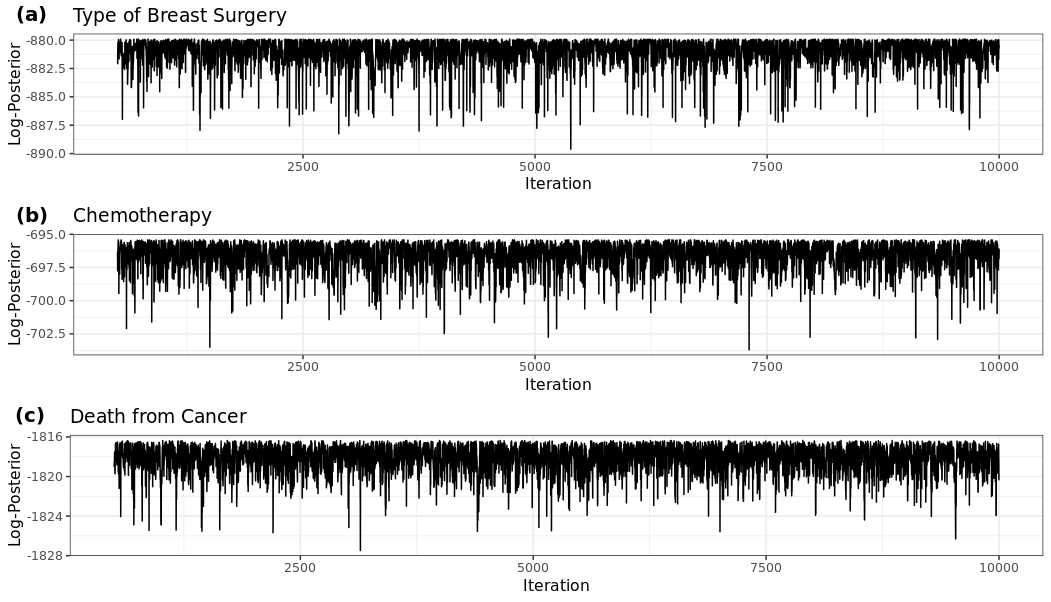}
    \caption{Traceplots for the log-posterior for TBS (a), CHT (b), and DFC (c). Burn-in iterations have been removed.}
    \label{fig:METABRIC-traces}
\end{figure}

The METABRIC dataset is \href{https://www.kaggle.com/datasets/raghadalharbi/breast-cancer-gene-expression-profiles-metabric}{publicly available}. We remove all missing values before analyzing the data. Age is dichotimized into decades and tumor size is dichotimized using quartiles. Recall that we implement Algorithm \ref{alg:edge-indicators} to infer the parent sets of the outcomes TBS (type of breast surgery), CHT (chemotherapy), and DFC (death from cancer). We set $\Ca(j)$ to be the union of the covariates and the previous outcomes in the stochastic order, so we have that $|\Ca(\tx{TBS})| = 9$, $|\Ca(\tx{CHT})| = 10$, and $|\Ca(\tx{DFC})| = 11$. Under the stochastic order, there are $2^9 + 2^{10} + 2^{11} = 3,584$ possible graphs in the DAG space.

The hyperparameters are $b_j=1$, $\rho_j = k_j+1$, $c_j=d_j=1$. $\bs t_j$ are initialized at $\bs t_j(x_j)=1$. We tune the stepsizes in a similar manner to Section \ref{sect:Markov-blanket-sims} in the main article: we begin by initializing $\bs \epsilon_j(x_j) = 0.1$, then increased or decreased based on the acceptance probabilities. We avoid very small or large values of the acceptance proportion. Once we have calibrated the stepsizes so that the acceptance probabilities are not extreme, we also look at the traceplots of $\bs t_j$ and $\log f(\bs t_j, \bs z_j \mid \bs n)$ to determine whether the chain is actually exploring the posterior distribution and if there are regions in which lots of rejections are occurring. For the three outcomes, the stepsizes are (0.175, 0.25), (0.095, 0.045), and (0.35, 0.2, 0.35); and with acceptance probabilities (0.552, 0.606), (0.629, 0.735), and (0.544, 0.655, 0.627). The traceplots for the log-posteriors of each outcome are displayed in Figure \ref{fig:METABRIC-traces}, and the ESS for the log-posteriors are 1947.235, 1791.852, and 1814.737. 
The HC and TABU algorithms are implemented using \texttt{bnlearn} and done sequentially. That is, we use the \texttt{blacklist} argument in \texttt{hc()} and \texttt{tabu()} to ensure the desired stochastic ordering. Lasso is implemented in the same way as in Section \ref{sect:Markov-blanket-sims} in the main article.

\section{Additional Experiment: Scoring Two Parent Sets} \label{section:structure-learning-sims}

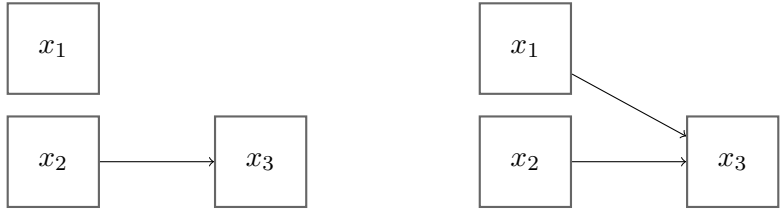
\begin{figure}[t]
    \centering
    \begin{tikzpicture}[
        node distance=1.5cm,
        squarednode/.style={rectangle, draw=black!60, thick, minimum size=12mm}
    ]

        \node[squarednode] (x1b) {$x_1$};
        \node[squarednode] (x2b) [below of=x1b] {$x_2$};
        \node[squarednode] (x3b) [right=1.5cm of x1b.east |- x2b] {$x_3$};

        \draw[->] (x2b) -- (x3b);
    \end{tikzpicture}
    \hspace{2cm}
    \begin{tikzpicture}[%
        node distance=1.5cm,
        squarednode/.style={rectangle, draw=black!60, thick, minimum size=12mm}
    ]
        \node[squarednode] (x1a) {$x_1$};
        \node[squarednode] (x2a) [below of=x1a] {$x_2$};
        \node[squarednode] (x3a) [right=1.5cm of x1a.east |- x2a] {$x_3$}; 

        \draw[->] (x1a) -- (x3a);
        \draw[->] (x2a) -- (x3a);
    \end{tikzpicture}
    \caption{Two possible DAGs, $G_1$ and $G_2$, respectively, for $p=3$ variables.}
    \label{fig:structure-learning-example}
\end{figure}

Next, we provide an additional example of using HiDDeN to evaluate two possible parent sets. We set $p=3$, $k_1=5$, $k_2=3$, and $k_3=2$, and set the two possible parent sets to be those displayed in Figure \ref{fig:structure-learning-example}. These are equivalent to setting $\Pa(3) = \lb 2 \rb$ and $\Pa(3) = \lb 1,2 \rb$, respectively. We vary $n \in \lb 50, 75, 100, 150  \rb $ and generate $100$ independent datasets from both DAGs for each sample size. In both cases, we simulate data according to $\Pr(x_1) = 1/k_1$, $\Pr(x_2) = 1/k_2$, and
\begin{gather*}
    \Pr(x_3=1 \mid x_2) \sim \tx{Unif}(0,1); \tx{ if } G_1 \tx{ is the true DAG}; \\
    \Pr(x_3=1 \mid x_1,x_2) \sim \tx{Unif}(0,1) ; \tx{ if } G_2 \tx{ is the true DAG}.
\end{gather*}
For each true DAG and replication, we compute $\widehat{\Pr}(G_{\tx{true}} \mid \bs n)$ via HiDDeN, as well as the BIC, AIC, and the BDE score for $G_1$ and $G_2$. The HiDDeN MCMC sampler with Algorithm \ref{alg:BPS} is ran for $20,000$ iterations with the first $200$ discarded as burn-in. The stepsize hyperameter is tuned using acceptance probabilities; $\bs \epsilon_3 = (0.5,0.5)$ when $G_1$ is the true DAG, and $\bs \epsilon_3 = (0.1,0.1)$ when $G_2$ is the true DAG. In either case, $\Pr(G_m)=1/2$. sTo evaluate the methods, we compute $\Pr(G_{\tx{true}} \mid \bs n)$ under HiDDeN and the differences in scores between the truth and alternative. For example, we set $\Delta \tx{BIC} = \tx{BIC}(G_{\tx{true}}) - \tx{BIC}(G_{\tx{false}})$, where $G_{\tx{false}}$ is the incorrect DAG, and we define this quantity analogously for all scores.

The results of our simulations are displayed in Table \ref{tab:structure-learning-sims}. To interpret the results, recall that HiDDeN, BIC, AIC, and BDE can be used for graph selection. For HiDDeN, we select the MAP, i.e. $G_{\tx{MAP}} = G_{\tx{true}}$ if $\Pr(G_{\tx{true}} \mid \bs n) > 1/2$, and $G_{\tx{MAP}} = G_{\tx{false}}$ otherwise. For the scores, e.g. BIC, we favor $G_{\tx{true}}$ when $\Delta \tx{BIC}>0$; this holds for $\Delta \tx{AIC}$ and $\Delta \tx{BDE}$ as well. When $G_1$ is the simulated network, all methods favor $G_1$. HiDDeN consistently allocates substantial posterior probability to $G_1$, which increases with $n$. When $G_{\tx{true}} = G_2$, the graph scores have difficulty in inferring the network structure. Both BIC and BDE on average favor $G_1$ (recall that this model has fewer parameters), though the scores have high variability in comparison to when $G_{\tx{true}} = G_1$. AIC fares better, with a positive average difference for all sample sizes, but also has erratic variability. This scenario has a sparser contingency table structure, as $\bs n_{\Pa(3),3}$ has dimension $5 \times 3 \times 2$, explaining the performance of BIC, AIC, and BDE. At the smallest sample size we consider, this amounts to just $50$ observations being spread out to $30$ cells. Notably, HiDDeN performs similarly in this challenging scenario to when $G_{\tx{true}} = G_1$, highly favoring $G_2$ for all sample sizes.

\begin{table}
    \centering
    \begin{tabular}{ccccc}
    \toprule
    \multicolumn{5}{c}{$G_{\tx{true}} = G_1$} \\
    \midrule
        $n$ & $\Pr(G_{\tx{true}} \mid \bs n)$  & $\Delta$ BIC & $\Delta$ AIC & $\Delta$ BDE \\
    \midrule
        50 & .813 (.247) & 17.319 (2.760) &  5.846 (2.760) & 14.051 (4.831) \\
        75 & .846 (.246) & 19.063 (2.931) &  5.158 (2.931) & 19.331 (7.014) \\
        100 & .925 (.155) & 21.427 (2.363) & 5.796 (2.363) & 23.916 (6.116) \\
        150 & .943 (.160) &  23.205 (2.899) & 5.141 (2.899) & 28.177 (6.430) \\
    \midrule
    \multicolumn{5}{c}{$G_{\tx{true}}=G_2$} \\
    \midrule
        $n$ & $\Pr(G_{\tx{true}} \mid \bs n)$  & $\Delta$ BIC & $\Delta$ AIC & $\Delta$ BDE \\
    \midrule
        50 & .818 (.296) & -8.774 (4.234) &  2.698 (4.234) & -4.020 (7.233) \\
        75 & .888 (.256) & -7.486 (4.807) &  6.419 (4.807) & -7.240 (8.223) \\
        100 & .908 (.238) & -4.999 (6.970) & 10.632 (6.970) & -6.812 (10.339) \\
        150 & .948 (.191) &  .099 (10.364) & 18.163 (10.364) & -4.421 (13.953) \\
    \bottomrule
    \end{tabular}
    \caption{Averages and standard deviations for the posterior probability of the true DAG under HiDDeN and the differences in BIC, AIC, and BDE. }
    \label{tab:structure-learning-sims}
\end{table}

\begin{table}
    \centering
    \begin{tabular}{ccccc}
        \toprule
        n & 50 & 75 & 100 & 150\\
        \midrule
        HiDDeN & 0.83 & 0.92 & 0.93 & 0.95 \\
        BDE & 0.27 & 0.19 & 0.25 & 0.34 \\
        \bottomrule
    \end{tabular}
    \caption{Proportion of replications in which HiDDeN and BDE selected $G_2$ when $G_{\tx{true}}=G_2$, defined as $\Pr(G_2 \mid \bs n) > 1/2$ and $\Delta \tx{BDE} >0$.}
    \label{tab:selected-DAGS}
\end{table}

Both our method and the BDE involve the Dirichlet distribution for graph selection, motivating a comparison between the two. Table \ref{tab:selected-DAGS} displays the proportion of the $100$ replications in which $G_2$ is favored in the scenario in which $G_{\tx{true}} = G_2$, defined as $\textbf{1}_{\Pr(G_2 \mid \bs n) > 1/2}$ for HiDDeN and $\textbf{1}_{\Delta \tx{BDE} >0}$ for BDE, where $\textbf{1}_{\{ \}}$ is the indicator function. BDE consistently favors the incorrect DAG, while HiDDeN prefers $G_2$ in more than $80\% $ of the replications for all sample sizes, with steadily increasing power. Table \ref{tab:selected-DAGS} provides further support for our methodology, as the BDE is constructed from the conjugate Dirichlet-multinomial model frequently used for Bayesian networks.

\section{Additional Example: the ALARM Network} \label{section:application} 

\begin{table}[t]
    \centering
    \begin{tabular}{ccccc}
    \toprule
        Name & Description & Class & Categories & $k_j$  \\
    \midrule
         LVF & Left ventricular failure & Diagnosis   & True, false &  2  \\
         HYP & Hypovolemia & Diagnosis &   True, false & 2   \\
         HIST & Patient history & Measurement  & True, false & 2   \\
         LVV & Left ventricular end-diastolic volume & Intermediate & Low, normal, high & 3   \\
         STKV & Stroke volume & Intermediate  & Low, normal, high & 3   \\
         PCWP & Pulmonary capillary wedge pressure & Measurement  & Low, normal, high & 3   \\
         CVP & Central venous pressure  & Measurement & Low, normal, high & 3   \\
    \bottomrule
    \end{tabular}
    \caption{Name, description, class, categories, and number of categories for the selected variables in the ALARM network.}
    \label{tab:alarm-variables}
\end{table}

\subsection{Background}

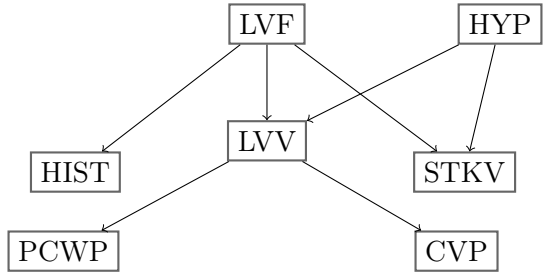
\begin{figure}[t]
    \centering
    \begin{tikzpicture}[roundnode/.style={circle, draw=black!60, minimum size=15mm}, squarednode/.style={rectangle, draw=black!60, thick, minimum size=5mm},box/.style = {draw,black,inner sep=10pt,rounded corners=5pt}, node distance = 2cm]
        % nodes
        \node[squarednode] (LVF) {LVF};
        \node[squarednode] (HYP) [right = of LVF] {HYP};
        \node[squarednode] (HIST) [below left = of LVF] {HIST};
        \node[squarednode] (LVV) [below = of LVF, yshift=10mm] {LVV};
        \node[squarednode] (STKV) [below right = of LVF] {STKV};
        \node[squarednode] (pcwp) [below left = of LVV, yshift=5mm] {PCWP};
        \node[squarednode] (cvp) [below right = of LVV, yshift=5mm] {CVP};
        % arrows
        \draw[->] (LVF)--(HIST);
        \draw[->] (LVF)--(LVV);
        \draw[->] (LVF)--(STKV);
        \draw[->] (HYP)--(LVV);
        \draw[->] (HYP)--(STKV);
        \draw[->] (LVV)--(pcwp);
        \draw[->] (LVV)--(cvp);
        % % boxes
    \end{tikzpicture}
    \caption{Network structure for a subset of variables in the ALARM dataset \citep{beinlich1989alarm}. }
    \label{fig:alarm-network}
\end{figure}

ALARM (A Logical Alarm Reduction Mechanism) is a Bayesian network designed using clinical knowledge to provide $8$ possible diagnoses to patients, with the ultimate goal of providing medical advice \citep{beinlich1989alarm}. There are $37$ total categorical variables in the network, including $8$ diagnoses, $16$ clinical measurements, and $13$ intermediate variables. The package \texttt{bnlearn} provides a set of $20000$ simulated observations from the ALARM network, and this dataset is frequently used to evaluate Bayesian network algorithms. In this paper, we focus on a subset of $p=7$ variables from the ALARM network including LVF, HYP, HIST, LVV, STKV, PCWP, and CVP. Figure \ref{fig:alarm-network} shows the known network structure of these variables, and Table \ref{tab:alarm-variables} gives their definitions. Variables either consist of true or false categories ($k_j=2$) or low, normal, or high categories ($k_j=3$); the selected subset consists of two diagnoses, three measurements, and two intermediate variables.

\subsection{Conditional Probabilities}

\begin{figure}
    \centering
    \includegraphics[scale=0.36]{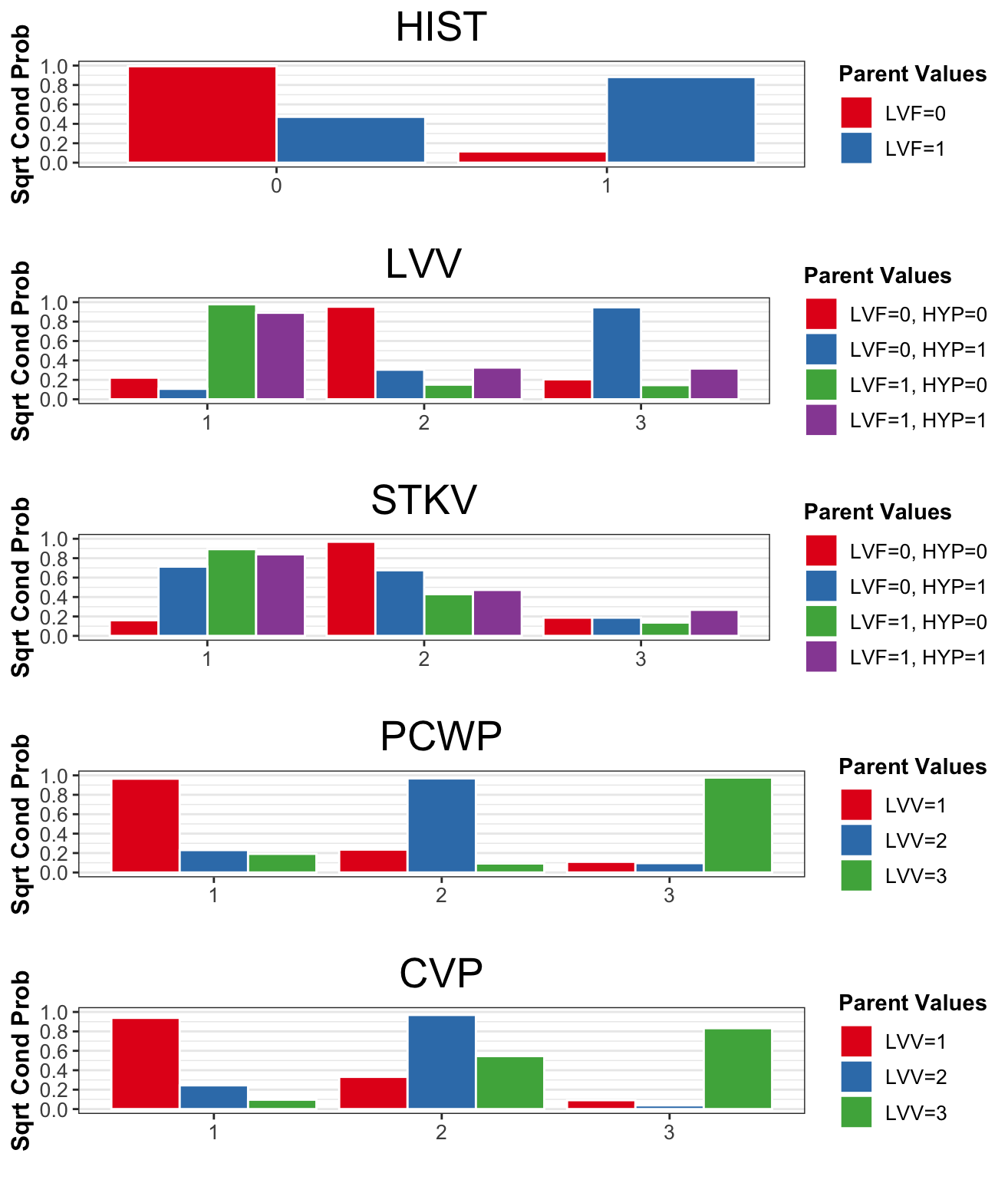}
    \caption{The values of $\sqrt{\widehat{\Pr}(\xnew_j \mid \bs x_{\Pa(j)},\bs n,G)}$ estimated from $n=200$ observations in the ALARM network. $\lb \tx{False}, \tx{True} \rb$ is coded as $\lb 0,1 \rb$ and $\lb \tx{Low}, \tx{Normal}, \tx{High} \rb$ is coded as $\lb 1,2,3 \rb$.  }
    \label{fig:alarm-conditional-probs}
\end{figure}

We use HiDDeN to infer the conditional probabilities of all non-roots: HIST, LVV, STKV, PCWP, and CVP. We take a random uniform sample of $n=200$ observations included in the dataset ($1\%$ of the data) to induce sparsity. For all variables, we run the MALA-within-Gibbs sampler (Algorithm \ref{alg:prior-means}) for $50,000$ iterations and set hyperparameters as $b_j=1$ and $\rho_j=k_j+1$. The MALA stepsizes are $(0.3,0.3)$, $(0.2,0.2,0.2)$, $(0.4, 0.4, 0.2)$, $(0.2,0.2,0.2)$, $(0.2,0.25,0.15)$ for HIST, LVV, STKV, and PCWP. The minimum and maximum ESS (taken across the conditional probabilities) for the variables are $(1805.76, 6630.32)$, $(3322.81, 6963.54)$, $(3156.89, 11156.64)$, $(4088.379, 6079.777)$, and $(3334.91, 6435.28)$, respectively. 

Figure \ref{fig:alarm-conditional-probs} plots the values of $\sqrt{\Pr(\xnew_j \mid \bs x_{\Pa(j)},\bs n,G)}$ for all $x_j$ and $\bs x_{\Pa(j)}$, where we take the square root for visualization purposes. First, given a patient has not experienced left ventricular failure, their probability of having no prior medical history is very high. Conversely, conditional on left ventricular failure, a patient is more likely to have medical history. Moreover, we can see that the absence of left ventricular failure and hypovolemia results in a higher probability of normal LVV. Additionally, the presence of hypovolemia and absence of left ventricular failure results in higher LVV. Finally, the presence of left ventricular failure is associated with low LVV. Both PCWP and CVP have direct relationships to the categories of LVV, though normal LVV values have non-negligible probability of leading to normal or high CVP. This relationship makes sense in the applied context because both PCWP and CVP are measurements of chamber pressure in the heart.

\subsection{Network Structure}

We next use HiDDeN to infer the network structure of the variables HIST, LVV, and STKV. Note from Figure \ref{fig:alarm-network} that these variables are children of LVF and/or HYP. We keep the $n=200$ observations from the previous section, and infer the parent sets in sequential order using Algorithm \ref{alg:BPS}. To ensure that the parent sets result in a DAG almost-surely, we evaluate all combinations of parent sets of the nodes in the previous sequence (excluding empty parent sets), imposing a stochastic order on the variables. For example, for LVV, we evaluate parent sets constructed from all possible sub-samples of $\lb \tx{LVF}, \tx{HYP}, \tx{HIST} \rb$, with the exception of the empty sample, $\emptyset$. Hence, $M_{\tx{HIST}} = 3$, $M_{\tx{LVV}} = 7$, and $M_{\tx{STKV}} = 15$. The candidate parent sets are all assigned equal prior probability. Each MCMC sampler is ran for $20000$ iterations, with the first $200$ removed as burn-in, and the stepsize hyperparameters are kept the same as the previous subsection.

For the three nodes, the parent sets with the maximum posterior probabilities (and their associated posterior probabilities) are $\lb \tx{LVF} \rb$ (0.825), $\lb \tx{LVF} ,\tx{HYP} \rb$ (0.961), and $\lb \tx{LVF} ,\tx{HYP} \rb$ (0.793), respectively; these are the exact parent sets in the ALARM network in Figure \ref{fig:alarm-network}. Hence, the MAP estimate of the parent sets accurately captures the true network. The only other parent set with positive probability for HIST is $\lb \tx{LVF},\tx{HYP} \rb$ (0.175), and $\lb \tx{HIST}, \tx{LVF}, \tx{HYP} \rb$ (0.039) has positive posterior probability as a parent set of LVV. There are $10$ parent sets for STKV with non-zero posterior probability, including $\lb \tx{HIST}, \tx{LVF}, \tx{HYP} \rb$ (0.172), the remaining parent sets have posterior probabilities totaling $0.035$. HiDDeN is excellent at inferring the individual edges originating at diagnostic variables (Table \ref{tab:alarm-edge-probs}): each true edge in Figure \ref{fig:alarm-network} has posterior probability greater than $0.993$, and the only edge not present in the true DAG has zero posterior mass.

\begin{table}
    \centering
    \begin{tabular}{ccc}
    \toprule
        \Pa(HIST) & LVF  & LVF, HYP \\
        \midrule
         $\Pr(\Pa(\tx{HIST}) \mid \bs n)$& 0.825 & 0.175 \\
    \bottomrule
    \end{tabular}

    \vspace{1mm}
    
    \centering
    \begin{tabular}{ccc}
    \toprule
        \Pa(LVV) & LVF,HYP  & HIST, LVF, HYP \\
        \midrule
         $\Pr(\Pa(\tx{LVV}) \mid \bs n)$& 0.961 & 0.039 \\
    \bottomrule
    \end{tabular}

    \vspace{1mm}

    \centering
    \begin{tabular}{ccccc}
    \toprule
        \Pa(STKV) & LVF,HYP  & HIST, LVF, HYP & LVV, LVF, HYP & HIST, LVF, HYP, LVV  \\
        \midrule
         $\Pr(\Pa(\tx{LVV}) \mid \bs n)$& 0.793 & 0.172 & 0.020 & 0.005 \\
    \bottomrule
    \end{tabular}
    
    \caption{The most visited parent sets in the MCMC sampler for HIST, LVV, and STVK, along with their posterior probabilities. The columns are ordered by posterior probability mass. $\Pa(\tx{STKV})$ visited $10$ total parent sets, so the rows of STKV in the above table do not sum to $1$. The remaining parent sets have posterior mass of $0.01$. }
    \label{tab:alarm-parent-probs}
\end{table}

\begin{table}
    \centering
    \footnotesize
    \begin{tabular}{ccccccc}
    \toprule
        Edge $e$ & (LVF,HIST) & (LVF,LVV)  & (LVF,STKV)  & (HYP,HIST)  & (HYP,LVV) & (HYP,STKV) \\
    \midrule
        $\widehat{\Pr}(e \mid \bs n)$ & 1 & 1 &  0.993 & 0  & 1  & 0.997 \\
    \bottomrule
    \end{tabular}
    \caption{Posterior probabilities for all directed edges originating from diagnosis variables. The only edge not present in Figure \ref{fig:alarm-network} is (HYP,HIST).}
    \label{tab:alarm-edge-probs}
\end{table}

\section{Maximum Marginal Likelihood Estimation}

\subsection{Convexity and Projected Gradient Descent}
 
In this section, we compute \textit{maximal marginal likelihood estimators} (MMLEs) for $\btildepi_j = \bs t_j/\beta_j$ for a \textit{fixed} $\beta_j>0$. Our results can be interpreted as an empirical Bayes approach \citep{berger1985statistical,bernardo1994bayesian}, a MAP estimation approach when $\btildepi_j \sim \tx{Dir}(1, \dots, 1)$ a priori, as well as an empirical technique for selecting the hyperparameters in the BDE graph score. We force the constraint that $\sum_{x_j=1}^{k_j} \btildepi_j(x_j) = 1$ and the negative log-likelihood for $\btildepi_j$ is
\begin{equation} \label{eq:objective-function}
    g_j(\bomega) = - \sum_{\bs x_{\Pa(j)}}\sum_{x_j=1}^{k_j} \lb  \log \Gamma( \bs n(\bs x_{\Pa(j)},x_j) + \beta_j \bomega(x_j) )- \log \Gamma(\beta_j \bomega(x_j)) \rb,
\end{equation}
where we are interested in minimizing $g_j$ over all $\bomega$ values in the probability simplex of dimension $k_j$, here denoted as $C_{k_j}$. Fixing $\beta_j$ results in the following property for $g_j$.
\begin{prop} \label{prop:strict-convexity}
   For any $\beta_j>0$, $g_j(\bomega)$ is strictly convex over $C_{k_j}$ if $\bs n_j(x_j)>0$ for all $x_j \in [k_j]$.
\end{prop}
Thus, there is a unique MMLE for all choices of $\beta_j$, derived from the optimization problem $\btildepi^{\tx{MMLE}}_j = \underset{\bomega \in C_{k_j}}{\tx{argmin }} g_j(\bs \omega)$. The proof of Proposition \ref{prop:strict-convexity}, displayed below, utilizes similar results to the proof of Theorem \ref{thm:log-concave}.

In practice, one can solve for $\btildepi^{\tx{MMLE}}_j$ across all nodes in $G$ and use projected gradient descent. The algorithm iteratively updates $\bomega$ by moving along the gradient of $g_j(\bomega)$: $\btildepi_j^{t+1} = \tx{proj}_{C_{k_j}}(\bomega^{t}-\eta_j \nabla g_j( \btildepi^{t}_j))$,
where $\eta_j>0$ and $\tx{proj}_{C_{k_j}}(\bs z)$ is the projection of any $\bs z \in \mathbb{R}^{k_j}$ onto $C_{k_j}$. One can project iterates onto the simplex using the technique in \cite{chen2011projection}. The gradient is evaluated via
\begin{equation} \label{eq:MMLE-gradient}
    \frac{\partial g}{\partial \bomega(x_j)} = - \beta_j \sum_{\bs x_{\Pa(j)}} \lb \psi( \bs n_{\Pa(j),j}(\bs x_{\Pa(j)},x_2) + \beta_j \bomega(x_j) )- \psi(\beta_j \bomega(x_j)) \rb,
\end{equation}
where $\psi(z)=\Gamma^\prime(z)/\Gamma(z)$ is the digamma function. Since $\lim_{z\to 0^{+}} \Gamma(y+z)/\Gamma(z) = \infty$ for all $y>0$, $\eta_j$ must be sufficiently small in order to avoid a projection in which $\bomega^{t+1}(x_j)=0$ for some $x_j \in [k_j]$. 

\subsection{Selecting $\beta_j$}
The predictive probability of transitioning from $\Pa(j)$ to any category in $[k_j]$ is
\begin{equation} \label{eq:shrinkage-estimator}
    \Pr(x_j^{\tx{new}}=x_j \mid \bs x_{\Pa(j)}^{\tx{new}}, \btildepi_j, \beta_j, \bs n_{\Pa(j),j} ) = \lambda_j \btildepi_j(x_j) + (1-\lambda_j) \widehat{\bs \pi}_{j \mid \Pa(j)}(x_j \mid \bs x_{\Pa(j)}),
\end{equation}
where $\widehat{\bs \pi}_{j \mid \Pa(j)}(x_j \mid \bs x_{\Pa(j)}) = \bs n_{\Pa(j),j}(\bs x_{\Pa(j)}, x_j)/\bs n_{\Pa(j)}(\bs x_{\Pa(j)})$ is the conditional cell probability MLE and $\lambda_j = \beta_j/\lb \beta_j + \bs n_{\Pa(j)}(\bs x_{\Pa(j)}) \rb$. This estimator is an extension of the James-Stein estimator for contingency table analysis \citep{hausser2009entropy} to the setting of directed graphical modeling. Hence, one can adapt commonly used Dirichlet hyperparameters associated with Jeffrey's prior ($\beta_j = k_j/2$), the Bayes-Laplace uniform prior ($\beta_j = k_j$), and Perks prior ($\beta_j = 1$), as in \cite{hausser2009entropy}. Observe that both Jeffrey's prior and the Bayes-Laplace uniform prior will result in more pronounced shrinkage towards $\btildepi_j$ for nodes with a large number of categories. This property may be appealing in this regime; higher $k_j$ will often result in more sparsity in $\bs n_{j, \Pa(j)}$. Other possible methods would be to estimate $\beta_j$ using optimization methods on the marginal likelihood, e.g. the Newton-Raphson algorithm for Dirichlet concentration parameters \citep{ronning1989maximum}, which is often employed for LDA \citep{blei2003latent}, though this involves fixing $\btildepi_j$. By alternating between these methods and the projected gradient descent, one can create estimates of $(\bs \beta_j, \btildepi_j)$.

\subsection{Proof of Proposition \ref{prop:strict-convexity}}
\begin{proof}
Clearly, for any $x_j \neq y_j \in [k_j]$, we have that
\begin{equation*}
    \frac{\partial^2 g_j}{\partial \bomega(x_j) \partial \bomega(y_j) } = \frac{\partial^2 g_j}{\partial \bomega(y_j) \partial \bomega(x_j) } =  0,
\end{equation*}
i.e., the Hessian is diagonal. Therefore, convexity is ensured so long as the diagonals of the Hessian are positive. We have that
\begin{equation} \label{eq:g-domega-second}
     \frac{\partial^2 g_j}{\partial \bomega(x_j)^2} = - \beta_j^2 \sum_{\bs x_{\Pa(j)}} \lb \psi^{(1)}( \bs n_{\Pa(j),j}(\bs x_{\Pa(j)},x_j) + \beta_j \bomega(x_j) )- \psi^{(1)}(\beta_j \bomega(x_j)) \rb
\end{equation}
for all $x_j$. As $\psi^{(1)}(z)$ is decreasing for all $z>0$, we have that $\frac{\partial^2 g_j}{\partial \bomega(x_j)^2} \geq 0$. Observe that \eqref{eq:g-domega-second} is positive when $\forall m_2 \in [k_2] \: \exists m_1 \in [k_1] \tx{ so that } \bs n_{\Pa(j),j}(\bs x_{\Pa(j)}, x_j) > 0$, which verifies Proposition \ref{prop:strict-convexity}.    
\end{proof}

\section{Structure Learning with Marginal Likelihood} \label{section:SM-marginal-likelihood}

\subsection{Marginal Likelihood}

This section focuses on both deriving and computing $ f(\bs n \mid \bs \beta, \bs \alpha, G)$, the marginal PMF of the contingency table after integrating out $\btildepi$, under the assumption that $\btildepi_j \sim \tx{Dir}(\alpha_j, \dots, \alpha_j)$. The marginal likelihood can be used to compute the Bayes Factor (BF) \citep{kass1995bayes} for multiple candidate DAGs.  Below, we show that the likelihood contribution is available in a closed form by summing across the discrete auxiliary variables introduced by \cite{teh2006sharing} for computation in the HDP.
\begin{theorem} \label{thm:marginal-likelihood}
    Let $f_j(G) =  f(\bs n_{j \mid \Pa(j)} \mid \bs n_{\Pa(j)}, \beta_j, \alpha_j, G)$ be the likelihood contribution of any non-root node in $G$. Then,
    \begin{equation} \label{eq:L_j(G)}
        f_j(G) =  H_j(G) \sum_{\bs w_{\Pa(j),j}} \beta_j^{w_j} \hspace{-5mm} \prod_{\bs x_{\Pa(j)}, x_j} \hspace{-3mm} s(\bs n_{\Pa(j),j}(\bs x_{\Pa(j)},x_j), \bs w_{\Pa(j),j}(\bs x_{\Pa(j)},x_j))|\times \frac{\prod_{x_j}\Gamma(\bs w_j(x_j) + \alpha_j)}{\Gamma(w_j + \alpha k_j)},
    \end{equation}
    where $\bs w_{\Pa(j),j} = (\bs w_{\Pa(j),j}(\bs x_{\Pa(j)},x_j))$ are  contingency tables such that $1 \leq \bs w_{\Pa(j),j}(\bs x_{\Pa(j)},x_j) \leq \bs n_{\Pa(j),j}(\bs x_{\Pa(j)},x_j)$ for all $\bs x_{\Pa(j)} \in \chi_{\Pa(j)}$ and $x_j \in [k_j]$, $w = \sum_{\bs x_{\Pa(j)}, x_j} \bs w_{\Pa(j),j}(\bs x_{\Pa(j)},x_j)$, $s(c,d)$ are the unsigned Stirling numbers of the first kind, and $H_j(G)$ is a constant depending on $G$, taking the form
    \begin{equation*}
        H_j(G) = \prod_{\bs x_{\Pa(j)}} \frac{\Gamma(\beta_j)}{\Gamma(\bs n_{\Pa(j)}(\bs x_{\Pa(j)}) + \beta_j)} \times \frac{\Gamma(k_j \alpha_j)}{\Gamma(\alpha_j)^{k_j}}.
    \end{equation*}
\end{theorem}
Theorem \ref{thm:marginal-likelihood} shows that we can compute the marginal likelihood of the HiDDeN model exactly by looping over the entries of $\bs n_{\Pa(j),j}$ in parallel for all $j$. This can be straightforward in some cases. For example, in the binary rooted tree, this can be implemented by applying four nested loops through the integers $\bs n_{j-1,j}(0,0)$, $\bs n_{j-1,j}(0,1)$, $\bs n_{j-1,j}(1,0)$, and $\bs n_{j-1,j}(1,1)$, since every node has no more than one parent. More general cases greatly increase the computation time of $f_j(G)$, motivating the more efficient methods that we propose in Section \ref{section:structure-learning}.

\subsection{Proof of Theorem \ref{thm:marginal-likelihood}}
\begin{proof}
    To show the result, we simplify notation. For the $j$th node, let $[k_1]$ denote the categories of its parents, $[k_2]$ the categories of $x_j$, and $\bs n$ the parent-child contingency table, that is, $\bs n_{j, \Pa(j)}$. Since the multinomial normalization constant does not depend on $G$, we will focus on the quantity
    \begin{equation*}
        \bs L(\bs n \mid \bs \pi) \propto \prod_{x_1} \prod_{x_2} \bs \pi(x_2 \mid x_1)^{\bs n(x_1,x_2)}.
    \end{equation*}
    First, we integrate out the conditional probabilities given $\btildepi_2$. We derive this expression in Section \ref{section:methods} of the main article, which is
    \begin{equation*}
        \mathcal L(\bs n \mid \btildepi_j) = \prod_{x_1}\frac{\Gamma(\beta)}{\Gamma(\bs n(x_1) + \beta)} \prod_{x_2} \frac{\Gamma(\bs n(x_1,x_2) + \beta \btildepi_2(x_2))}{\Gamma(\beta \btildepi_2(x_2))}.
    \end{equation*}
    We expand the ratios of Gamma functions as in the Gibbs sampler from Section \ref{section:posterior} to obtain
    \begin{equation*}
        \frac{\Gamma(\bs n(x_1,x_2) + \beta \btildepi_2(x_2))}{\Gamma(\beta \btildepi_2(x_2))} = \sum_{ \bs w(x_1,x_2) = 1    }^{\bs n(x_1,x_2)} s( \bs n(x_1,x_2), \bs w(x_1,x_2) ) \beta^{ \bs w(x_1,x_2)} \btildepi_2(x_2)^{\bs w(x_1,x_2)},
    \end{equation*}
    where $s(n, w)$ are the unsigned Sitrling numbers of the first kind, and note that $s(0,w)=0$. This implies that
    \begin{equation} \label{eq:SM-ML-normalizing}
        \prod_{x_1,x_2}\frac{\Gamma(\bs n(x_1,x_2) + \beta \btildepi_2(x_2))}{\Gamma(\beta \btildepi_2(x_2))} = \sum_{\bs w}  \prod_{x_1,x_2} s( \bs n(x_1,x_2), \bs w(x_1,x_2) ) \beta^{ \bs w(x_1,x_2)} \btildepi_2(x_2)^{\bs w(x_1,x_2)},
    \end{equation}
    where $\bs w$ sums overall all contingency tables whose entries are bounded by those in $\bs n$. That is, $\bs w = (\bs w(x_1,x_2))_{x_1,x_2}$ such that $1 \leq \bs w(x_1,x_2) \leq \bs n(x_1,x_2)$ for all $x_1 \in [k_1], x_2 \in [k_2]$. 

    Now, we must integrate over $\btildepi_2$, which has a $\tx{Dir}(\alpha, \dots, \alpha)$ distribution. This means solving the integral
    \begin{gather}
        \frac{\Gamma(k_2 \alpha)}{\Gamma(\alpha)^{k_2}} \int \sum_{\bs w}  \prod_{x_1,x_2} s( \bs n(x_1,x_2), \bs w(x_1,x_2) ) \beta^{ \bs w(x_1,x_2)} \btildepi_2(x_2)^{\bs w(x_1,x_2) + (\alpha -1)/k_1} \nonumber \diff \btildepi_2 \nonumber \\
        = \frac{\Gamma(k_2 \alpha)}{\Gamma(\alpha)^{k_2}} \sum_{\bs w} \beta^{w} \times \prod_{x_1,x_2} s(\bs n(x_1,x_2), \bs w(x_1,x_2)) \times \int \prod_{x_2} \btildepi_2(x_2)^{\bs w(x_2) + \alpha - 1} \nonumber \\
        = \frac{\Gamma(k_2 \alpha)}{\Gamma(\alpha)^{k_2}} \sum_{\bs w} \beta^{w} \times  \prod_{x_1,x_2} s(\bs n(x_1,x_2), \bs w(x_1,x_2)) \times \frac{\prod_{x_1} \Gamma(\bs w(x_2) + \alpha)}{\Gamma(w + k_2 \alpha_2)}. \label{eq:SM-ML-final-sum}
    \end{gather}
    Combining \eqref{eq:SM-ML-final-sum} with the normalizing constant in \eqref{eq:SM-ML-normalizing} completes the proof.

\end{proof}

\end{document}